\documentclass[twocolumn,prb,aps,citeautoscript,superscriptaddress,longbibliography]{revtex4-2}

\usepackage{graphicx}
\usepackage{amsmath}
\usepackage{amssymb}
\usepackage{marvosym}
\usepackage{txfonts}
\usepackage{bm}
\usepackage{xcolor}
\usepackage{color}
\usepackage{ulem}
\usepackage{makecell}
\usepackage{multirow}
\usepackage[colorlinks=true,urlcolor=blue,anchorcolor=blue,linkcolor=blue,citecolor=blue,breaklinks=true]{hyperref}

\newcommand{\lsco}{La$_{2-x}$Sr$_x$CuO$_4$}
\newcommand{\lscod}{La$_{2-x}$Sr$_x$CuO$_{4+\delta}$}
\newcommand{\lcco}{La$_{2-x}$Ca$_x$CuO$_4$}
\newcommand{\lco}{La$_{2}$CuO$_{4.12}$}

\newcommand{\ybco}[1]{YBa$_2$Cu$_3$O$_{#1}$}
\newcommand{\bscco}{Bi$_2$Sr$_2$CaCu$_2$O$_{8 + \delta}$}
\newcommand{\cecoin}{CeCoIn$_5$}
\newcommand{\sro}{Sr$_2$RuO$_4$}
\newcommand{\lamb}{$\lambda$}

\newcommand{\tc}{\ensuremath{T_c}}

\newcommand{\swave}{\mbox{$s$-wave}}
\newcommand{\dwave}{\mbox{$d$-wave}}

\def\={&=&}

\definecolor{forestgreen(web)}{rgb}{0.13, 0.55, 0.13}

\definecolor{LSCORed}{RGB}{214,39,40}

\definecolor{TlBlue}{RGB}{31, 119, 180}

\definecolor{ForestGreen}{rgb}{0.13, 0.55, 0.13}

\definecolor{Orange}{RGB}{255, 100, 0}
\definecolor{Cobaltb}{RGB}{0, 71, 171}

\definecolor{Purple}{RGB}{128,0,128}

\begin{document}

\normalem


\title{Beyond Homes scaling: disorder, the Planckian bound and a new universality}

\author{D.~M.~Broun}
\affiliation{Department of Physics, Simon Fraser University, Burnaby, BC, V5A~1S6, Canada}
\author{Vivek Mishra}
\affiliation{School of Engineering and Science, Indian Institute of Technology Madras Zanzibar Campus, Bweleo, Zanzibar 71215, Tanzania}
\affiliation{Department of Physics, University of Florida, Gainesville FL 32611}
\author{J.~S.~Dodge}
\affiliation{Department of Physics, Simon Fraser University, Burnaby, BC, V5A~1S6, Canada}
\author{P.~J. Hirschfeld}
\affiliation{Department of Physics, University of Florida, Gainesville FL 32611}

\begin{abstract}
Beginning with high-$T_c$ cuprate materials, it has been observed that many superconductors exhibit so-called ``Homes scaling'', in which the zero-temperature superfluid density, $\rho_{s0}$, is proportional to the product of the normal-state \mbox{dc conductivity} and the superconducting transition temperature, $\sigma_\mathrm{dc} T_c$.  For conventional, \swave\ superconductors, such scaling has been shown to be a natural consequence of elastic-scattering disorder, not only in the extreme dirty limit but across a broad range of scattering parameters.  Here we show that when an analogous  calculation is carried out for elastic scattering in  \dwave\ superconductors, a stark contrast emerges, with $\rho_{s0} \propto \left(\sigma_\mathrm{dc} T_c \right)^2$ in the dirty limit, in apparent violation of Homes scaling.  Within a simple approximate Migdal--Eliashberg treatment of inelastic scattering, we show how  Homes scaling is recovered. The normal-state behavior of near optimally doped cuprates is dominated by inelastic scattering, but significant deviations from Homes scaling occur for disorder-dominated cuprate systems, such as underdoped \ybco{6.333} and overdoped \lsco, and in very clean materials with little inelastic scattering, such as Sr$_2$RuO$_4$.  We present a revised analysis where both axes of the original Homes scaling plot are normalized by the Drude plasma weight, $\omega_{p,D}^2$, and show that new universal scaling emerges, in which the superfluid fractions of dirty \swave\ and dirty \dwave\ superconductors coalesce to a single point at which normal-state scattering is occurring at the Planckian bound.  The combined result is a new tool for classifying superconductors in
terms of order parameter symmetry, as well as scattering strength and character.
Although our model starts from a Fermi-liquid assumption it describes underdoped cuprates surprisingly well.
\end{abstract}

\maketitle{}

\section{Introduction}
\label{sec:introduction}

Since condensed matter physics deals with a huge variety of  materials, universal relations highlighting commonalities among disparate systems  are often extremely valuable to deduce fundamental laws of nature.  Well-known examples are given by the universal ratios that occur in the BCS theory of superconductivity \cite{bcs}, relating properties of paired Fermi systems with interactions and degeneracy temperatures varying over many orders of magnitude, or scaling relations  that reflect the irrelevance of details of a system near its critical point \cite{Wilson.1975}.  One scaling relation that appears extremely important but whose origin has proved elusive is so-called ``Homes scaling", proposed initially  by Dordevic et al.~\cite{Dordevic2002} and Homes et al.~\cite{Homes:2004p354} as a way to highlight common behavior of several families of cuprate superconductors. ``Homes' Law" states that the zero-temperature superfluid density $\rho_{s0} \equiv \rho_s(T\!=\!0) \propto \sigma_{\mathrm{dc}} T_c$, where $\sigma_{\mathrm {dc}}$ is the dc conductivity and $T_c$ is the critical temperature.  While several theoretical proposals \cite{Zaanen2004,Tallon2006,BasovChubukov2011,Erdmenger2012} have invoked exotic aspects of cuprate physics such as non-Fermi liquid behavior to explain Homes scaling,  others have derived similar scaling based on models of granularity \cite{Imry2012} or pointlike disorder in dirty \swave\ superconductors~\cite{Homes.2005,Kogan:2013eh}.  The inadequacy of any single one of these explanations was underlined by the extension of Homes scaling to conventional electron--phonon and organic superconductors \cite{Dordevic:2013ep}, i.e., including systems with both $s$- and \dwave\ pairing, with significantly different degrees of inhomogeneity and disorder.

We believe that the role of disorder in Homes scaling deserves further attention, since at first glance impurities should cause very different effects in $s$- and \dwave\ superconductors: nonmagnetic disorder has virtually no effect on $T_c$ in \swave\ systems  whereas even weak disorder breaks \dwave\ pairs.  We therefore begin a re-examination of Homes scaling by calculating the superfluid density, $T_c$ and conductivity for a simple \dwave\ model in the presence of disorder, and show that it produces a fundamentally different scaling in the dirty limit. This raises the question of why cuprates obey Homes scaling at all. 

To address this question, we consider the effects of inelastic scattering via a simple Migdal--Eliashberg model with \swave\ or \dwave\ pairing \cite{Millis:1988},  employing a strong-scattering boson spectrum that has been used to  roughly characterize  optimally doped cuprates \cite{Schachinger.1997}. When the characteristic boson frequency, $\Omega_\mathrm{B}$, is several times $T_c$, one obtains a normal state scattering rate \mbox{$\hbar \Gamma \sim k_B T_c$} --- i.e.,  a scattering rate at the Planckian limit. We then show that when both axes of the original Homes scaling plot are rescaled by the Drude plasma weight, $\omega_{p,D}^2$, a new universal scaling emerges, in which the superfluid fractions of dirty \swave\ and dirty \dwave\ superconductors coalesce to a single point at which normal-state scattering is occurring at the Planckian bound.  This restores Homes scaling for \dwave\ superconductors dominated by strong inelastic scattering (e.g., optimally doped cuprates and the heavy fermion superconductor CeCoIn$_5$), but retains the stark dichotomy between \swave\ and \dwave\ superconductors in cases where the normal-state inelastic scattering is weak (e.g., heavily underdoped \ybco{6.333} and strongly overdoped \lsco).  The revised scaling plot is well supported by data on a wide range of conventional and unconventional superconductors, and suggests that the prevalence of Planckian-limit scattering in many different materials  \cite{Hartnoll.2022} underlies the ubiquity of the original Homes scaling.

\begin{figure}[t]
    \centering
        \includegraphics[width=1.0\columnwidth,scale=1.0]{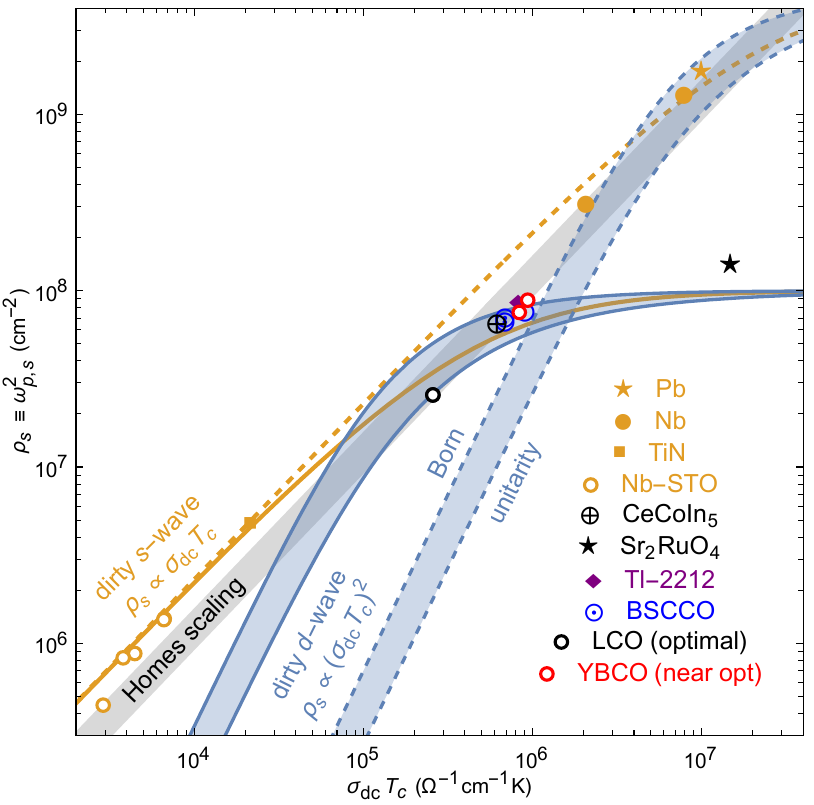}
    \caption{Homes scaling (gray line) shown alongside theory curves for pointlike elastic scattering disorder, revealing different dirty-limit power-laws for \swave\ and \dwave\ superconductors.  Elastic scattering curves have been plotted for two choices of Drude plasma frequency: \mbox{$\omega_{p,D} = 10^4$~cm$^{-1}$} (solid lines) characteristic of typical cuprates; and \mbox{$\omega_{p,D} = 7 \times 10^4$~cm$^{-1}$} (dashed lines) characteristic of an elemental metal such as Nb. In the \dwave\ case, results depend on the scattering phase shift of the impurities and are bounded by the Born and unitarity limits.  Also plotted is a selection of data for various superconductors, used throughout the paper and summarized in Table~\ref{tab:materials}.
}
    \label{fig:standard_Homes_scaling}
\end{figure}

\begin{figure}[t]
    \includegraphics[width=1.0\columnwidth,scale=1.0]{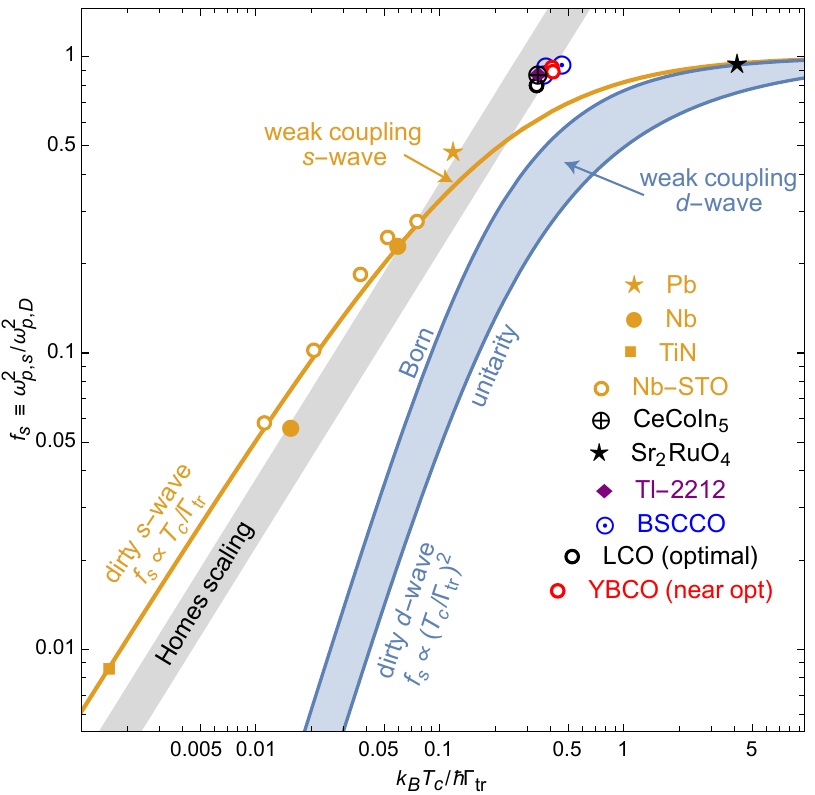}
    \caption{The revised superfluid scaling plot factors out the material-specific Drude weight, $\omega_{p,D}^2$, to reveal the underlying universal behavior, expressed in terms of the dimensionless quantities \mbox{$f_s \equiv \omega_{p,s}^2/\omega_{p,D}^2$} and $k_B T_c/\hbar \Gamma_\mathrm{tr}$.  Rescaled data from Fig.~\ref{fig:standard_Homes_scaling} are plotted, along with the \swave\ and \dwave\ curves for elastic scattering disorder. The rescaled Homes line is shown for reference.}
    \label{fig:ModifiedHomesScaling}
\end{figure}

Here we present our main results by way of summarizing the logical flow of the paper.
Details of derivations and discussion of data are postponed until later sections.  First,   in Sec.~\ref{sec:elastic_scattering}, we  review expected Homes scaling in systems with elastic scattering only.  The result in the case of isotropic \swave\ superconductors  \cite{Homes.2005,Kogan:2013eh}, that $\rho_{s0} \propto \sigma_\mathrm{dc} T_c$ in the dirty limit, is recovered.  However, a similar calculation for the $d$-wave case, which seems  not to have been presented heretofore, yields a different scaling relation, $\rho_{s0}\propto (\sigma_\mathrm{dc}T_c)^2$.  This very different behavior in the dirty limit arises of course from the fact that potential scatterers break pairs in $d$-wave superconductors, but may also be related more intuitively to the conductivity spectral weight lost in the low-$T$ superconducting state, as will be discussed later.  These very different scaling behaviors are shown
superposed on a familiar Homes scaling plot in Fig.~\ref{fig:standard_Homes_scaling}, including experimental data for various superconductors, as detailed below and in Table~\ref{tab:materials}.  It is easily seen that over much of the plot the data follow the results for the dirty $s$-wave case rather than the dirty $d$-wave case.      
To reveal what we believe to be a more fundamental universality, we replace the traditional Homes scaling plot with a revised superfluid scaling plot where the key material-specific property --- the Drude weight --- has been scaled out of both $\rho_{s0}$ and $\sigma_\mathrm{dc} T_c$, yielding a plot of the superfluid {\it fraction} $f_s = \rho_{s0}/\omega_{p,D}^2$ (here $\rho_{s0}$ is expressed in plasma frequency units, as in the original Homes paper, and $\omega_{p,D}^2$ is the Drude weight) vs.\ the critical temperature normalized to the transport scattering rate, $T_c/\Gamma_\mathrm{tr}$.  The data are again plotted in Fig.~\ref{fig:ModifiedHomesScaling} and in the dirty limit (low $T_c/\Gamma_\mathrm{tr}$) show a clear separation between \swave\ and \dwave\ cases. 

It is conspicuous, however, that many of the original, near optimally doped cuprates have been mapped in this procedure to a small cluster of points in Fig.~\ref{fig:ModifiedHomesScaling}, and furthermore do not lie on either of the elastic-scattering curves.      We therefore continue in Section \ref{sec:inelastic} and analyze the effects of inelastic scattering on Homes scaling for $s$- and \dwave\ superconductors.  Of course, we cannot present a complete theory of inelastic processes in all superconductors, particularly in cuprates.  However, we show that a simple Eliashberg-like approach describing scattering of fermions from a simple, model bosonic spectrum
does a remarkable job of describing the revised superfluid scaling behavior of a very large variety of superconducting materials and, in its strong-coupling limits, provides an explanation for the nearly universal scaling behavior observed in the Planckian regime. These results, which constitute 
the main findings of our paper, are given in Fig.~\ref{fig:modified_Homes_scaling_inelastic}. Superconductors are shown to exhibit the revised superfluid scaling not only according  to their order parameter symmetry, but also depending on whether they are dominated by elastic or inelastic scattering processes.

In Sec.~\ref{sec:experimental_data}, we detail the methods used to obtain the experimental quantities $\rho_{s0}$, $\omega_{p,D}^2$  and $\Gamma_\mathrm{tr}$ and give sources for the data, which are summarized in \mbox{Table.~\ref{tab:materials}}.

In Sec.~\ref{sec:optics_pair_breaking}, we use optical spectra to provide an intuitive perspective of the key differences between strong inelastic scattering, dirty \swave\ superconductivity, and dirty \dwave\ superconductivity, and show how these result in the different scaling regimes observed in the revised superfluid scaling plot.

In Sec.~\ref{sec:Discussion}, we summarize how the process of factoring out the Drude weight, $\omega_{p,D}^2$, reveals the true universal superfluid scaling, resulting in four distinct scaling regimes: Planckian, superclean, dirty \swave, and dirty \dwave, along with the material examples that fall into each category.  We discuss the implications of our results, including the possibility that a much broader range of Homes-scaling behavior can be understood within a Fermi-liquid framework than might previously have been imagined.  

Finally, we conclude in Sec.~\ref{sec:Conclusions} by proposing ways to ways to test and extend our analysis with new experimental data, emphasizing the importance of techniques that simultaneously probe the Drude weight, superfluid density and transport scattering rate.  We also note the utility of the revised superfluid scaling plot as a powerful tool for classifying new materials, giving immediate insight into disorder level, order-parameter symmetry, and strength of coupling to the bosonic fluctuations responsible for superconducting pairing, inelastic scattering and mass renormalization.

The paper has been structured so that a reader wishing to skip the technical details of elastic and inelastic scattering in superconductors may omit Secs.~\ref{sec:elastic_scattering} and \ref{sec:inelastic} on a first reading.

\section{Elastic scattering}

\label{sec:elastic_scattering}

\subsection{General dirty BCS formalism}

The Nambu-space Green’s function for a dirty \dwave\ superconductor can be written as \cite{Hirschfeld:1994}
\begin{equation}
G(i \omega_n; \phi,\xi)=- \frac{i \tilde\omega_n \tau_0 + \tilde\Delta_n(\phi) \tau_1 + \xi \tau_3}{\tilde\omega_n^2 + \tilde\Delta_n^2(\phi) + \xi^2}\;.
\end{equation}
Here $\tilde\Delta_n(\phi) = \tilde\Delta(i \omega_n;\phi)$ is the renormalized gap, in general varying with angle $\phi$ around the Fermi surface.  $\tilde \omega_n$ is a renormalized fermionic Matsubara frequency, $\xi$ is the single-particle dispersion relative to the Fermi level, and $\tau_i$ are the Pauli particle--hole matrices.

\begin{figure}[t]
    \centering
        \includegraphics[width=1.0\columnwidth,scale=1.0]{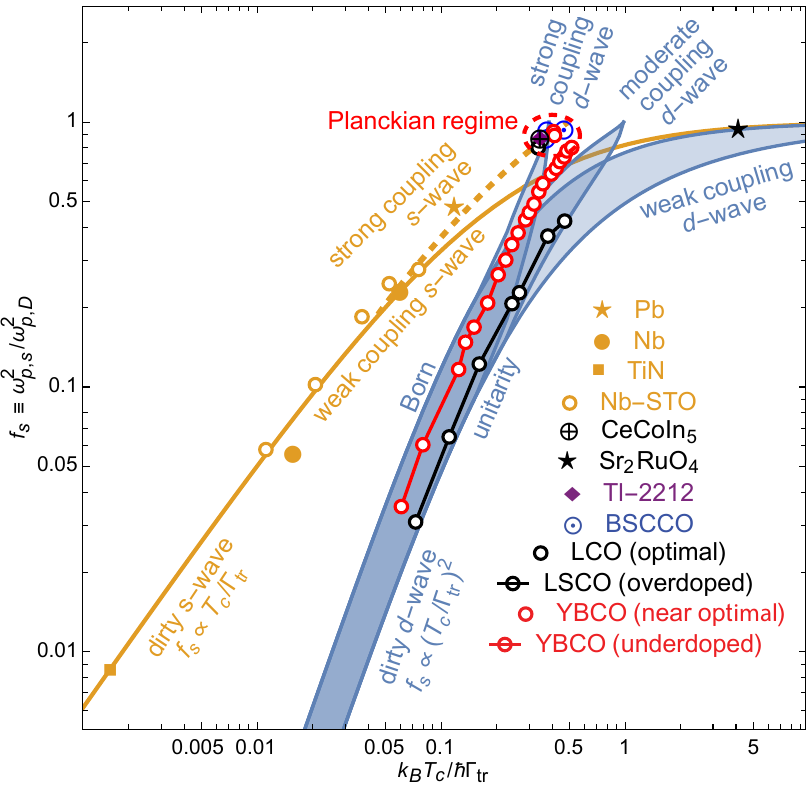}
    \caption{Revised superfluid scaling, including the effects of inelastic scattering and disorder.  For the \dwave\ case, three regimes of inelastic scattering are shown: strong coupling ($\Omega_B = 3.5 T_{c0}$); intermediate coupling ($\Omega_B = 8 T_{c0}$); and weak coupling (BCS limit). The \dwave\ behavior is delimited in each case by the Born and unitarity limits for point scatterers.  In the \swave\ case, only strong coupling and BCS limits are shown.  The near optimally cuprates and the heavy fermion superconductor \cecoin\ are dominated by inelastic scattering and cluster tightly in the Planckian regime.  Radiating out along one line are the dirty \swave\ superconductors.  Dirty \dwave\ superconductors follow a different scaling line, as typified by overdoped \lsco\ and underdoped \ybco{6.333}. Superclean \sro\ exists in a class of its own. }
    \label{fig:modified_Homes_scaling_inelastic}
\end{figure}

The BCS gap equation is 
\begin{equation}
\Delta(\phi) =  - 2\pi T \sum_{\omega_n>0}^{\Omega_c} \left\langle N_0 V_p(\phi,\phi') \frac{\tilde \Delta_n(\phi')}{\sqrt{\tilde \omega_n^2 + \tilde \Delta_n^2(\phi')}}\right\rangle_\mathrm{FS}\;,
\label{eq:gap_equation}
\end{equation}
where $N_0$ is the single-spin density of states at the Fermi level, $V_p(\phi,\phi')$ is the pairing interaction, $\Omega_c$ is a high frequency cut off, and $\langle...\rangle_\mathrm{FS}$ denotes an angle average over the Fermi surface.

For point-scattering disorder, the Matsubara frequency and gap are renormalized according to \cite{LPGorkov:1985,KUeda:1985,Schmitt-Rink:1986,Hirschfeld:1986ii,PROHAMMER:1991p557,Hirschfeld:1994}
\begin{align}
\tilde{\omega}_n &= \omega_n + \Gamma \frac{g_{0,n}}{c^2+g_{0,n}^2+g_{1,n}^2},\label{eq:renorm_Matsubara}\\
\tilde{\Delta}_n(\phi) &= \Delta(\phi) + \Gamma \frac{g_{1,n}}{c^2+g_{0,n}^2+g_{1,n}^2},
\label{eq:renorm_gap}
\end{align}
where the scattering-rate parameter $\Gamma = n_\mathrm{imp}/(\pi N_0)$, $n_\mathrm{imp}$ is the density of impurities, the cotangent of the scattering phase shift is $c = 1/(\pi N_0 V_\mathrm{imp})$ and $V_\mathrm{imp}$ is the strength of the pointlike impurity potential.  The integrated diagonal Green's function, averaged over disorder, is
\begin{equation}
g_{0,n} = \left \langle \frac{\tilde\omega_n}{\sqrt{\tilde\omega_n^2+ \tilde\Delta_n^2(\phi)}} \right\rangle_\mathrm{FS}\;.
\end{equation}
The integrated, disorder-averaged, off-diagonal Green's function is
\begin{equation}
g_{1,n} = \left \langle \frac{\tilde\Delta_n(\phi)}{\sqrt{\tilde\omega_n^2+ \tilde\Delta_n^2(\phi)}} \right\rangle_\mathrm{FS}\;.
\end{equation}
In the normal state, $g_0 = 1$ and $g_1 = 0$, so that $\tilde\omega_n = \omega_n + \Gamma_N$, where the normal-state scattering rate $\Gamma_N = \Gamma/(1 + c^2)$.

The transition temperature, $T_c$, is obtained by solving the linearized gap equation in the presence of disorder:
\begin{eqnarray}
\Delta(\phi) &=&  2\pi T \sum_{\omega_n>0}^{\Omega_c} \left\langle N_0 V(\phi,\phi') \frac{\tilde \Delta_n(\phi')}{\tilde \omega_n}\right\rangle_\mathrm{FS}\;.
\end{eqnarray}
In the Homes-law literature, which draws heavily from optics, superfluid density is usually presented in plasma-frequency units as $\omega_{p,s}^2 = c^2/\lambda_L^2$, where $\lambda_L$ is the familiar London penetration depth and $c = 1/\sqrt{\mu_0 \epsilon_0}$ is the speed of light.  At finite temperatures, the superfluid density (for current flowing in the $x$ direction) is given by \cite{NAM:1967p640}
\begin{equation}
    \rho_s \equiv \omega_{p,s}^2  = \frac{2 N_0 e^2}{\epsilon_0} 2 \pi T \sum_{\omega_n >0}^\infty \left\langle v_{F,x}^2(\phi) \frac{\tilde \Delta_n^2(\phi)}{\left(\tilde \omega_n^2 + \tilde \Delta_n^2(\phi) \right)^{3/2}}\right\rangle_\mathrm{FS}\;.
\end{equation}
In the limit of an isotropic Fermi surface --- i.e., when the Fermi velocity, $v_F(\phi)$, is independent of angle --- the superfluid density reduces to
\begin{equation}
    \rho_s = \omega_{p}^2\times2 \pi T \sum_{\omega_n >0}^\infty \left\langle \frac{\tilde \Delta_n^2(\phi)}{\left(\tilde \omega_n^2 + \tilde \Delta_n^2(\phi) \right)^{3/2}}\right\rangle_\mathrm{FS}\;,
    \label{eq:superfluid_density_isotropic}
\end{equation}
where $\omega_{p}^2 = N_0 e^2 v_F^2/\epsilon_0$ is the plasma frequency of the conduction electrons. For simplicity, we will assume an isotropic Fermi surface (but not an isotropic gap) in the remainder of the paper, without too much loss of generality.

At zero temperature, the Matsubara sum in Eq.~\ref{eq:superfluid_density_isotropic} becomes an integral:
\begin{equation}
    \rho_{s0} = \omega_{p}^2 \times \int_0^\infty d\omega \left\langle \frac{\tilde \Delta^2(\phi)}{\left(\tilde \omega^2 + \tilde \Delta^2(\phi) \right)^{3/2}}\right\rangle_\mathrm{FS}\;,
    \label{eq:zero_temperature_superfuid}
\end{equation}
 where it is understood that $\tilde \Delta(\phi)$ in general depends implicitly on 
 frequency $\omega$.

 Finally, the dc conductivity at $T_c$ is 
\begin{equation}
    \sigma_\mathrm{dc} = \frac{\epsilon_0 \omega_p^2}{\Gamma_\mathrm{tr}} = \frac{\epsilon_0 \omega_p^2}{2 \Gamma_N}\;,
\end{equation}
where, for point scatterers, the transport relaxation rate is \mbox{$\Gamma_\mathrm{tr} = 2 \Gamma_N$}.

\subsection{Dirty s-wave}

The relationship between Homes scaling and dirty \swave\ superconductivity has previously been reported \cite{Homes.2005,Kogan:2013eh}.  In the isotropic \swave\ case, $\tilde \Delta_n(\phi)= \tilde \Delta_n$. The integrated, disorder-averaged Green's functions defined above then have the property that \mbox{$g_{0,n}^2 + g_{1,n}^2 =1$}.  This makes the renormalization equations for $\tilde \omega_n$ and $\tilde \Delta_n$ take on a similar form to one another:
\begin{align}
    \tilde \omega_n & = \omega_n + \Gamma_N \frac{\tilde \omega_n}{\sqrt{\tilde \omega_n^2 + \tilde \Delta_n^2}}\\
    \tilde \Delta_n & = \Delta + \Gamma_N \frac{\tilde \Delta_n}{\sqrt{\tilde \omega_n^2 + \tilde \Delta_n^2}}\;,
\end{align}
with the result that the ratio $\tilde \omega_n/\tilde \Delta_n = \omega_n/\Delta$ is not renormalized by disorder scattering.  With isotropic gap and isotropic pairing interaction $V_p = -V$, the \swave\ gap equation
\begin{equation}
\Delta =  2\pi T N_0 V \!\!\sum_{\omega_n>0}^{\Omega_c} \! \frac{\tilde \Delta_n}{\sqrt{\tilde \omega_n^2 + \tilde \Delta_n^2}} = 2\pi T N_0 V \!\!\sum_{\omega_n>0}^{\Omega_c} \! \frac{\Delta}{\sqrt{\omega_n^2 + \Delta^2}}
\end{equation}
depends only on the ratio $\tilde \omega_n/\tilde \Delta_n$, and is therefore unchanged from its clean-limit form.  This leads to Anderson's theorem \cite{Anderson:1959p2774}, with $T_c$ and $\Delta$ protected from nonmagnetic elastic disorder.  In particular, the zero-temperature gap ratio, $ \Delta_0/k_B T_c = 1.76$, takes its clean-limit, universal value.

The zero-temperature superfluid density, Eq.~\ref{eq:zero_temperature_superfuid}, can similarly be recast in terms of unrenormalized quantities, but is not protected from disorder:
\begin{align}
    \rho_{s0} &= \omega_{p}^2 \int_0^\infty d\omega  \frac{\tilde \Delta^2}{\left(\tilde \omega^2 + \tilde \Delta^2 \right)^{3/2}}\\
    &= \omega_{p}^2 \int_0^\infty d\omega  \frac{\Delta_0^2}{\omega^2 + \Delta_0^2}\frac{1}{\sqrt{\omega^2 + \Delta_0^2}+\Gamma_N}\\
    & = \omega_{p}^2 \frac{\Delta_0}{\Gamma_N}\left(\frac{\pi}{2} - \cosh^{-1}\left(\Gamma_N/\Delta_0\right)/\sqrt{\Gamma_N^2/\Delta_0^2 - 1} \right)\;.
\end{align}
In the clean limit ($\Gamma_N \ll \Delta_0$)
\begin{equation}
    \rho_{s0} \approx \omega_{p}^2\left(1 - \frac{\pi \Gamma_N}{4\Delta_0}\right)\;.
\end{equation}
In the dirty limit ($\Gamma_N \gg \Delta_0$)
\begin{align}
    \rho_{s0} \approx \frac{\pi}{2}\omega_{p}^2 \frac{\Delta_0}{\hbar \Gamma_N} = 1.76 \pi \frac{k_B}{\epsilon_0 \hbar} \sigma_\mathrm{dc} T_c\;,
    \label{eq:Homes_scaling_swave}
\end{align}
revealing the dirty \swave\ Homes scaling reported previously  \cite{Homes.2005,Kogan:2013eh}.  The full crossover from dirty to clean is plotted in Fig.~\ref{fig:standard_Homes_scaling}.

\subsection{Dirty d-wave}

We now explore the connection between Homes scaling and dirty \dwave\ superconductivity.  For concreteness and simplicity, we continue to assume an isotropic, circular Fermi surface, on which the \dwave\ gap takes the form $\Delta(\phi) = \Delta \cos 2 \phi$.  (We reserve the notation $\Delta_0$ to refer to the zero-temperature gap maximum of the disordered system.)  The \dwave\ gap averages to zero around the Fermi surface, so that the integrated off-diagonal Green's function vanishes identically: $g_{1,n} = 0$.   As a result, there is no explicit renormalization of the gap in Eq.~\ref{eq:renorm_gap}, and no dependence of $\Delta$ on Matsubara index in our BCS approximation.  
The effects of disorder enter solely through the renormalized Matsubara frequency \cite{Hirschfeld.1993zw}:
\begin{equation}
    \tilde \omega_n = \omega_n + \Gamma \frac{g_{0,n}}{c^2 + g_{0,n}^2}\;.
    \label{eq:renorm_Matsubara_dwave}
\end{equation}
To aid in numerical evaluation, the various angle averages in this section are carried out explicitly on the isotropic, circular Fermi surface and expressed in terms of complete elliptic integrals of the first and second kind,
\begin{align}
K(m)&= \int_{0}^{\frac{\pi}{2}} \frac{d\theta}{\sqrt{1-m \sin^2\theta}}\;, \\
E(m)&= \int_{0}^{\frac{\pi}{2}}  \sqrt{1-m \sin^2\theta}\;d\theta\;.
\end{align}
For the \dwave\ superconductor,
\begin{equation}
g_{0,n} = \left \langle \frac{\tilde\omega_n}{\sqrt{\tilde\omega_n^2+\Delta^2 \cos^2 2\phi}} \right\rangle_\mathrm{FS} = f_1\left(\frac{\Delta^2}{\tilde \omega_n^2}\right),
\end{equation}
where
\begin{equation}
    f_1(x) = \frac{2}{\pi} \frac{1}{\sqrt{1+x}}K\left(\frac{x}{1+x}\right)\;.
\end{equation}
To produce a \dwave\ gap on the circular Fermi suface, we employ a separable pairing interaction
\begin{equation}
V_p(\phi,\phi')= -V \cos 2\phi \cos 2\phi'.
\end{equation}
The finite-temperature gap equation is
\begin{equation}
    \Delta =  N_0 V 2\pi T \sum_{\omega_n>0}^{\Omega_c} \left\langle \frac{\Delta \cos^2 2\phi}{\sqrt{\tilde \omega_n^2 + \tilde \Delta^2\cos^2 2\phi}}\right\rangle_\mathrm{FS}\;.
\end{equation}
The zero-temperature \dwave\ gap equation is
\begin{align}
\Delta_0 &= N_0 V \int_0^{\Omega_c}\!\!\! d\omega \left\langle\frac{\Delta_0 \cos^2 2\phi}{\sqrt{\tilde \omega^2+\Delta_0^2 \cos^2 2\phi}}\right\rangle_\mathrm{FS}\\
& = N_0 V \int_0^{\Omega_c}\!\!\! d\omega \; f_2\left(\frac{\Delta_0^2}{\tilde \omega^2}\right)\;,
\end{align}
where
\begin{equation}
    f_2(x) = \frac{2}{\pi} \sqrt{1+1/x}\left[E\left(\frac{x}{1+x}\right) - \frac{1}{1+x}K\left(\frac{x}{1+x}\right)\right]\;.
\end{equation}
The zero-temperature superfluid density is  
\begin{align}
    \rho_{s0} & = \omega_{p}^2\times\!\!\int_0^\infty \!\!\! d \omega \left\langle \frac{ \Delta_0^2\cos^2 2\phi}{\left(\tilde \omega^2 +  \Delta_0^2\cos^2 2\phi \right)^{3/2}}\right\rangle_\mathrm{FS}\\
    & = \omega_{p}^2\times\!\!\int_0^\infty \!\!\! d \omega \;\frac{1}{\tilde \omega} \;f_3\left(\frac{\Delta_0^2}{\tilde \omega^2}\right)\;,
\end{align}
where
\begin{equation}
    f_3(x) = \frac{2}{\pi} \frac{1}{\sqrt{1+x}}\left[K\left(\frac{x}{1+x}\right) - E\left(\frac{x}{1+x}\right)\right]\;.
\end{equation}
In the zero-temperature expressions, it is useful to change integration variable from $\omega$ to $\tilde \omega$ by replacing the lower bound in the integrals by $\tilde \omega_0 \equiv \tilde \omega(\omega = 0)$, from Eq.~\ref{eq:renorm_Matsubara_dwave}, and by differentiating the expression for renormalized Matsubara frequency, Eq.~\ref{eq:renorm_Matsubara}, with respect to $\tilde \omega$, from which we obtain
\begin{equation}
    d\omega = d\tilde\omega \left[ 1-\Gamma \frac{c^2-g_0^2}{\left(c^2+g_0^2\right)^2} g_0^\prime \right]\;,
\end{equation}
where $g_0 = g_0(\tilde \omega)$, $g_0' = dg_0/d\tilde \omega$, and $\Gamma = \Gamma_N(1+c^2)$.  That is, 
\begin{align}
\frac{1}{N_0 V} &= \int_{\tilde \omega_0}^{\Omega_c}\!\!\! d\tilde \omega \left\langle\frac{\cos^2 2 \phi}{\sqrt{\tilde \omega^2+\Delta_0^2 \cos^2 2\phi}}\right\rangle_\mathrm{FS} \left[ 1-\Gamma \frac{c^2-g_0^2}{\left(c^2+g_0^2\right)^2} g_0^\prime \right]\\
& = \int_{\tilde \omega_0}^{\Omega_c}\!\!\! d\tilde \omega \; \frac{1}{\Delta_0}f_2\left(\frac{\Delta_0^2}{\tilde \omega^2}\right) \left[ 1-\Gamma \frac{c^2-g_0^2}{\left(c^2+g_0^2\right)^2} g_0^\prime \right]\;,
\end{align}
and
\begin{align}
    \rho_{s0} & = \omega_{p}^2\times\!\!\int_{\tilde \omega_0}^\infty\!\!\! d\tilde \omega \left\langle \frac{ \Delta_0^2\cos^2 2\phi}{\left(\tilde \omega^2 \!+\!  \Delta_0^2\cos^2 2\phi \right)^{3/2}}\right\rangle_\mathrm{\!FS}\left[ 1-\Gamma \frac{c^2-g_0^2}{\left(c^2\!+\!g_0^2\right)^2} g_0^\prime \right]\\
    & = \omega_{p}^2\times\!\!\int_{\tilde \omega_0}^\infty\!\!\! d\tilde \omega \;\frac{1}{\tilde \omega} \;f_3\left(\frac{\Delta_0^2}{\tilde \omega^2}\right)\left[ 1-\Gamma \frac{c^2-g_0^2}{\left(c^2+g_0^2\right)^2} g_0^\prime \right]\;.\label{eq:rhos0_dwave}
\end{align}

The transition temperature is determined by solving the linearized gap equation to obtain the  Abrikosov--Gor'kov result
\begin{eqnarray}
\ln \left( \frac{T_c}{T_{c0}}\right) &=& \psi\left( \frac{1}{2}\right) - \psi\left( \frac{1}{2} + \frac{\Gamma_N}{2\pi T_{c}}\right)\;,
\label{eq:tc}
\end{eqnarray}
where $\psi(x)$ is the digamma function and the clean-limit transition temperature is
\begin{equation}
    T_{c0}=\frac{2e^\gamma}{\pi}\Omega_c \exp(-2/N_0 V)\;.
\end{equation}
Here $\gamma$ is the Euler–Mascheroni constant and we have assumed $T_{c0},\Gamma_N \ll \Omega_c$. The clean-limit energy gap is
\begin{equation}
\Delta_{00} = 2\pi e^{-\gamma-\frac{1}{2}} T_{c0} \approx 2.14 T_{c0}\;.
\end{equation}
In the weak-disorder limit ($\Gamma_N \ll T_{c0}$),
\begin{equation}
T_c \approx T_{c0} \left( 1- \frac{\pi}{4} \frac{\Gamma_N }{T_{c0}}\right)\;.
\end{equation}
The superconducting transition temperature is driven to zero at a critical scattering rate 
\begin{equation}
   \Gamma_\mathrm{cr} = \pi T_{c0}/(2e^\gamma) = \Omega_c \exp(-2/N_0 V)\;.
   \label{eq:critical_scattering_rate}
\end{equation}
 Near $\Gamma_\mathrm{cr}$, where $T_c \ll \Gamma_N$, Eq.~\ref{eq:tc} can be approximated by 
\begin{eqnarray}
\ln \left( \frac{T_c}{T_{c0}}\right) &\approx& -\ln \left(4 e^\gamma \right)- \ln \left(  \frac{\Gamma_N}{2\pi T_{c}} \right)-  \frac{1}{24}\left(\frac{2\pi T_{c}}{\Gamma_N}\right)^2;
\end{eqnarray}
leading to the result 
\begin{equation}
T_c \approx \Gamma_N \frac{\sqrt{6}}{\pi} \sqrt{\ln \left(\frac{\Gamma_\mathrm{cr}}{\Gamma_N}\right)} \approx \frac{\sqrt{6}}{\pi}\Gamma_\mathrm{cr} \sqrt{1-\frac{\Gamma_N}{\Gamma_\mathrm{cr}}}\;.
\label{eq:dirty_tc}
\end{equation}
In the asymptotic dirty regime, $\Delta_0 \to 0$ as $\Gamma_N \to \Gamma_\mathrm{cr}$.  In this limit, $\Delta_0 \ll \tilde \omega$, so all quantities can be expanded in the small parameter $\Delta_0/\tilde \omega$.  We start with the integrated Green's function $g_0(\tilde \omega)$:
\begin{align}
g_0(\tilde \omega) & = \left \langle \frac{\tilde\omega}{\sqrt{\tilde\omega^2+\Delta_0^2 \cos^2 2\phi}} \right\rangle_\mathrm{FS} \\
 & = \left \langle \left(1 + \frac{\Delta_0^2}{\tilde \omega^2} \cos^2 2\phi \right)^{-1/2}\right\rangle_\mathrm{FS}\\
  & \approx \left \langle 1 - \frac{\Delta_0^2}{2 \tilde \omega^2} \cos^2 2\phi \right\rangle_\mathrm{FS}\\
  & = 1 - \frac{\Delta_0^2}{4 \tilde \omega^2}\;.
  \label{eq:g0_expansion}
\end{align}
Its derivative is
\begin{equation}
    g_0'(\tilde \omega) = \frac{d g_0}{d \tilde \omega} \approx \frac{\Delta_0^2}{2 \tilde \omega^3}\;.
\end{equation} 
Substituting the expansion Eq.~\ref{eq:g0_expansion} into the renormalization equation for $\tilde \omega$ (Eq.~\ref{eq:renorm_Matsubara_dwave}) gives
\begin{equation}
    \tilde \omega \approx \omega + \Gamma_N\left(1 + \frac{\Delta_0^2}{4 \tilde \omega^2}\frac{1-c^2}{1+c^2}\right)\;,
\end{equation}
from which we obtain
\begin{equation}
    \tilde \omega_0 \approx \Gamma_N\left(1 + \frac{\Delta_0^2}{4 \Gamma_N^2}\frac{1-c^2}{1+c^2}\right)\;.
    \label{eq:omega_0}
\end{equation}
The kernel of the superfluid density is
\begin{equation}
    \left\langle \frac{ \Delta_0^2\cos^2 2\phi}{\left(\tilde \omega^2 \!+\!  \Delta_0^2\cos^2 2\phi \right)^{3/2}}\right\rangle_\mathrm{FS} \approx \frac{\Delta_0^2}{2 \tilde \omega^3}\;,
\end{equation}
so that the zero-temperature, dirty-limit superfluid density  from Eq.~(\ref{eq:rhos0_dwave}) is
\begin{align}
    \rho_{s0} & \approx \omega_{p}^2 \times \!\!\int_{\tilde \omega_0}^\infty \!\!\!d\tilde \omega \frac{\Delta_0^2}{2 \tilde \omega^3} \left[ 1-\Gamma \frac{c^2-g_0^2}{\left(c^2+g_0^2\right)^2} g_0^\prime \right]\\
    & \approx \omega_{p}^2 \times \!\!\int_{\tilde \omega_0}^\infty \!\!\!d\tilde \omega \frac{\Delta_0^2}{2 \tilde \omega^3}\\
    & \approx \omega_{p}^2 \frac{\Delta_0^2}{\Gamma_N^2}\;.
    \label{eq:dirty_dwave_superfluid}
\end{align}
This is the key result for the dirty \dwave\ superconductor, revealing that $\rho_{s0} \sim (\Delta_0/\Gamma_N)^2 \sim (T_c/\Gamma_N)^2$, distinct from the \swave\ case in which $\rho_{s0} \sim \Delta_0/\Gamma_N \sim T_c/\Gamma_N$.  To establish a direct relationship to $T_c$ we need to solve the zero-temperature gap equation in the dirty limit.  The kernel of the gap equation is
\begin{align}
    \left\langle\frac{\cos^2 2\phi}{\sqrt{\tilde \omega^2+\Delta_0^2 \cos^2 2\phi}}\right\rangle_\mathrm{FS} & = \frac{1}{\tilde \omega} \left\langle\cos^2 2\phi \left(1 + \frac{\Delta_0^2}{\tilde \omega^2} \cos^2 2\phi\right)^{-1/2}\right\rangle_\mathrm{FS}\\
    & \approx \frac{1}{2 \tilde \omega}\left(1 -\frac{3\Delta_0^2}{8 \tilde \omega^2}  \right)\;,
\end{align}
so that the dirty limit gap equation becomes
\begin{align}
    \frac{2}{N_0 V} & \approx \int_{\tilde \omega_0}^{\Omega_c}\!\!\! d\tilde \omega \left(\frac{1}{\tilde \omega} - \frac{3\Delta_0^2}{8 \tilde \omega^3} \right) \left[ 1-\Gamma \frac{c^2-g_0^2}{\left(c^2+g_0^2\right)^2} g_0^\prime \right]\\
    & \approx \int_{\tilde \omega_0}^{\Omega_c}\!\!\! d\tilde \omega \left[\frac{1}{\tilde \omega} - \Delta_0^2\left( \frac{3}{8\tilde\omega^3} + \frac{(c^2-1)\Gamma_N}{2(c^2+1)\tilde\omega^4}\right)\right]
\end{align}
We integrate the second term from $\tilde\omega=\Gamma_N$ to $\infty$, and for the first term we use the full expression for $\tilde \omega_0$, Eq. \eqref{eq:omega_0}. This gives
\begin{equation}
\frac{2}{N_0 V} \approx  \ln \left( \frac{\Omega_c}{\Gamma_N \left( 1+ \frac{\Delta_0^2(1-c^2)}{4\Gamma_N^2(1+c^2)}\right)}\right)  - \frac{\Delta_0^2 (1+17c^2)}{48\Gamma_N^2 (1+c^2)}\;.
\end{equation}
Using $2/N_0V = \ln(\Omega_c/\Gamma_\mathrm{cr})$, from Eq.~\ref{eq:critical_scattering_rate}, we have
\begin{align}
\ln \left( \frac{\Omega_c}{\Gamma_\mathrm{cr}} \right) &\approx  \ln \left( \frac{\Omega_c}{\Gamma_{N}} \right)  -\frac{\Delta_0^2(13+5c^2)}{48\Gamma_N^2 (1+c^2)} \\
\Delta_0 &\approx  4\sqrt{\frac{3(1+c^2)}{13+5c^2}}\Gamma_N \sqrt{\ln \left( \frac{\Gamma_\mathrm{cr}}{\Gamma_N}\right)}\;.
\end{align}
Comparing this with the asymptotic form of $T_c$ in the dirty limit (Eq.~\ref{eq:dirty_tc}) we obtain
\begin{equation}
    \frac{\Delta_0}{k_B T_c} \approx  4\pi\sqrt{\frac{(1+c^2)}{2(13+5c^2)}}\;.
\end{equation}
The gap ratio $\Delta_0/k_B T_c$ is 2.46 in the unitarity limit $(c = 0)$ and 3.97 in the Born limit $(c \to \infty)$.  The fact that the gap ratio in a \dwave\ superconductor depends on  scattering phase shift is the reason that the disorder-dominated behavior in Figs.~\ref{fig:ModifiedHomesScaling}  and \ref{fig:modified_Homes_scaling_inelastic} spreads out into broad bands rather than following a single curve, as in the \swave\ case.

Bringing these results together, we have for the asymptotic dirty \dwave\ superfluid density
\begin{align}
    \rho_{s0}^d & \approx  8 \pi^2 \frac{1+c^2}{13+5c^2} \omega_{p}^2 \left(\frac{k_B T_c}{\hbar \Gamma_N}\right)^2\\
    & =  32 \pi^2 \frac{1+c^2}{13+5c^2} \frac{k_B^2}{\epsilon_0^2 \hbar^2\omega_{p}^2} \left(\sigma_\mathrm{dc} T_c\right)^2\;,
    \label{eq:Homes_scaling_dwave}
\end{align}
exhibiting the quadratic dependence of $\rho_s$ on $\sigma_\mathrm{dc} T_c$.  This is to be contrasted with the dirty \swave\ scaling from Eq.~\ref{eq:Homes_scaling_swave},
\begin{equation}
    \rho_{s0}^s \approx 1.76 \pi \frac{k_B}{\epsilon_0 \hbar} \sigma_\mathrm{dc} T_c\;,
\end{equation}
in which the power law follows the standard linear form of Homes' law.

\section{Inelastic scattering}
\label{sec:inelastic}

As a minimal description of a strong-coupling superconductor with inelastic scattering, we employ a Migdal--Eliashberg model with a boson spectrum that has been used to roughly characterize optimally doped cupates \cite{Schachinger.1997}.  While no one would claim that this provides a complete theory of these materials, the advantage of the Migdal--Eliashberg approach is that it allows the effects of inelastic scattering, superconducting pairing,  interaction-induced renormalization of Drude weight, and disorder to be treated on an equal footing --- precisely the ingredients required for an analysis of the effects of inelastic scattering on  Homes scaling.

\subsection{d-wave Migdal--Eliashberg theory}

\label{sec:d_wave_Eliashberg}

Following Ref.~\cite{Schachinger.1997}, for the \dwave\ case the boson spectrum is assumed to contain both an \swave\ component and a separable \dwave\ component:
\begin{equation}
    I^2 F(\Omega,\phi,\phi^\prime) = I^2 F(\Omega)\left(1 + g \sqrt{2}\cos(2 \phi) \sqrt{2} \cos(2 \phi^\prime)\right)\;.
    \label{eq:boson_spectral_density}
\end{equation}
Here $g$ is a parameter that controls the relative strength of the coupling to $\tilde \omega$ and $\tilde \Delta$ channels.  We assume that the frequency dependence of the boson spectrum is the same in the $\tilde \omega$ and $\tilde \Delta$ channels and takes the form
\begin{equation}
    I^2F(\Omega) = I^2 \frac{\Omega/\Omega_B}{1 + \left(\Omega/\Omega_B\right)^2}\;.
    \label{eq:strong_coupling_spectrum}
\end{equation}
As in Ref.~\onlinecite{Schachinger.1997}, the characteristic boson energy is taken to be \mbox{$\Omega_B = 350~\mathrm{K} \approx 30$~meV}, and an upper cutoff  $\Omega_\mathrm{max} = 4650~\mathrm{K} \approx 400$~meV is applied to the spectrum.  With $g = 0.8$, the coupling strength $I^2$ can readily be tuned to give clean-limit \mbox{$T_c = 100$~K}, with $I^2 \approx 1.3$. Deep within the superconducting state, an additional low-frequency cutoff, $\omega_c = 210$~K, is used to capture the the gapping of the boson spectrum by the onset of superconductivity.  For a spectrum of the form of Eq.~\ref{eq:strong_coupling_spectrum}, the mass-renormalization parameter is
\begin{equation}
    \lambda = \int_0^\infty \frac{2 I^2 F(\Omega) d \Omega}{\Omega} = \pi I^2\;,
\end{equation}
and the characteristic frequency scale is 
\begin{equation}
    \Omega_\mathrm{log} = \exp\left(+\frac{2}{\lambda} \int_0^\infty \ln(\Omega) \frac{I^2 F(\Omega)}{\Omega} d \Omega \right) = \Omega_B\;.
\end{equation}
The parameters identified in Ref.~\onlinecite{Schachinger.1997} as corresponding to optimally doped cuprates, and adopted here, correspond to $\lambda \approx 4$ and a strong-coupling ratio $T_c/\Omega_\mathrm{log} \approx 0.3.$

Using similar notation to Sec.~\ref{sec:elastic_scattering}, and including a $t$-matrix term for point-scattering disorder, the imaginary-axis self-energy equations are

\begin{widetext}

\begin{equation}
\begin{split}
    \tilde \omega_n & = \omega_n + \Gamma \frac{g_{0,n}}{c^2 + g_{0,n}^2} + \pi T \sum_{m = -\infty}^\infty \lambda(m-n) \left \langle \frac{\tilde \omega_m}{\sqrt{\tilde \omega_m^2 + \tilde \Delta_m^2 \cos^2(2 \phi)}} \right \rangle_\mathrm{FS}\;,
\end{split}
\label{eq:renorm_Matsubara_Eliashberg}
\end{equation}
\begin{align}
         \tilde \Delta(i \omega_n; \phi) = \tilde \Delta_n  \cos(2 \phi) & = \pi T g\sum_{m = -\infty}^\infty \lambda(m-n) \sqrt{2} \cos(2 \phi) \left \langle  \frac{\sqrt{2} \cos(2 \phi')\tilde \Delta_m\cos(2 \phi')}{\sqrt{\tilde \omega_m^2 + \tilde \Delta_m^2\cos^2(2 \phi')}} \right \rangle_\mathrm{FS}\\
        & \Rightarrow \tilde\Delta_n = 2  \pi T g\sum_{m = -\infty}^\infty \lambda(m-n) \left \langle \frac{\tilde \Delta_m\cos^2(2 \phi)}{\sqrt{\tilde \omega_m^2 + \tilde \Delta_m^2\cos^2(2 \phi)}} \right \rangle_\mathrm{FS}\;.
\label{eq:renorm_gap_Eliashberg}
\end{align}

\end{widetext}
Here the interaction term is
\begin{equation}
    \lambda(m-n) = 2\int_0^\infty d \Omega \frac{\Omega\,I^2 F(\Omega)}{\Omega^2 + (\omega_m - \omega_n)^2}\;.
\end{equation}
For the strong-coupling spectrum, Eq.~\ref{eq:strong_coupling_spectrum}, $\lambda(m-n)$ can  be evaluated in closed form, including the effects of the spectral cutoffs $\omega_c$ and $\Omega_\mathrm{max}$:
\begin{equation}
    \lambda(m-n) = 2\,I^2 \!\!\int_{\omega_c}^{\Omega_\mathrm{max}} \!\!\!\!\! \frac{\Omega }{\Omega^2 + \left(\omega_m\!-\!\omega_n \right)^2} \frac{\Omega/\Omega_B}{1 + \left(\Omega/\Omega_B\right)^2} d\Omega
\end{equation}
Expressed in terms of the frequency difference $\nu \equiv \omega_m - \omega_n$ this evaluates to
\begin{multline}
    \lambda(\nu) = 2 I^2 \frac{\Omega_B}{ \Omega_B^2 - \nu^2}\Bigg[\Omega_B\left(\arctan\tfrac{\Omega_\mathrm{max}}{\Omega_B} - \arctan\tfrac{\omega_c}{\Omega_B}\right)\\- \nu\left(\arctan\tfrac{\Omega_\mathrm{max}}{\nu} - \arctan \tfrac{\omega_c}{\nu}\right) \Bigg]\;.
\end{multline}
In the absence of cutoffs, $\lambda(\nu) \to \pi I^2 \Omega_B/(\Omega_B+\nu)$.

At $T_c$, the renormalization equations can be linearized by setting $\tilde \Delta_m = 0$ in the denominators of Eqs.~\ref{eq:renorm_Matsubara_Eliashberg} and \ref{eq:renorm_gap_Eliashberg}:
\begin{equation}
    \tilde \omega_n = \omega_n + \Gamma_N + \pi T_c \!\!\!\sum_{m = -\infty}^\infty\!\!\! \lambda(m\!-\!n) \frac{\tilde \omega_m}{|\tilde \omega_m|}\label{eq:renorm_Matsubara_Tc}\;,
\end{equation}
\begin{equation}
         \tilde \Delta_n  = \pi T_c g\!\!\!\sum_{m = -\infty}^\infty\!\!\! \lambda(m\!-\!n) \frac{\tilde \Delta_m}{|\tilde \omega_m|}\label{eq:renorm_gap_Tc}\;.
\end{equation}

The  superfluid density is given by
\begin{align}
    \rho_s & = \omega_p^2 \times \pi T \! \sum_{n = -\infty}^\infty \Bigg \langle \frac{\tilde \Delta_n^2(\phi)}{\left(\tilde \omega_n^2 + \tilde \Delta_n^2(\phi)^2 \right)^{3/2}}\Bigg \rangle_\phi\\
    & = \omega_p^2 \times  2 \pi T \!\sum_{n >0}^\infty \Bigg \langle \frac{\tilde \Delta_n^2 \cos^2(2\phi)}{\left(\tilde \omega_n^2+ \tilde \Delta_n^2 \cos^2(2\phi) \right)^{3/2}}\Bigg \rangle_\phi\;.
\end{align}
Here $\omega_p$ is the plasma frequency corresponding to the \emph{full} conduction electron spectral weight.  One benefit of the Migdal--Eliashberg approach is that it explicitly captures the renormalization of Drude weight that occurs as bosonic fluctuations shift optical weight to higher frequencies.  This is the process usually thought of as interaction-induced mass enhancement.  A beautiful (and extreme) version can be seen in the rearrangement of the optical conductivity spectrum of the heavy fermion superconductor \cecoin, which takes place as Kondo physics emerges below room temperature \cite{Singley.2002}. As an operational definition of the Drude weight corresponding to a given boson spectrum, we define $\omega_{p,D}^2$ to be the renormalized zero-temperature superfluid density in the limit of zero disorder. 

Calculations of the Migdal--Eliashberg gap and self-energy have been carried out numerically at a temperature of $T_c/20$, with a Matsubara cut-off equal to $10 \Omega_\mathrm{max} = 465 T_{c0}$, and then extrapolated to obtain the zero-temperature superfluid density.  By the time $T_c$ is small enough that the calculations are no longer tractable, we are well into the dirty regime in which elastic scattering dominates and the exact asymptotic results from Sec.~\ref{sec:elastic_scattering} apply.

\subsection{s-wave Migdal--Eliashberg theory}
\label{sec:s_wave_Eliashberg}

The \swave\ calculation proceeds in a similar way to the \dwave\ case, but with an isotropic gap and boson spectrum, resulting in:
\begin{equation}
\begin{split}
    \tilde \omega_n & = \omega_n + \Gamma_N \frac{\tilde\omega_m}{\sqrt{\tilde \omega_m^2\!+\! \tilde \Delta_m^2}} + \pi T \!\!\sum_{m = -\infty}^\infty\!\!\! \lambda(m\!-\!n) \frac{\tilde\omega_m}{\sqrt{\tilde \omega_m^2\! +\! \tilde \Delta_m^2}}\;,
\end{split}
\end{equation}
\begin{equation}
         \tilde \Delta_n  = \Gamma_N \frac{\tilde\Delta_m}{\sqrt{\tilde \omega_m^2 + \tilde \Delta_m^2}} +  \pi T g\sum_{m = -\infty}^\infty \lambda(m-n) \frac{\tilde\Delta_m}{\sqrt{\tilde \omega_m^2 + \tilde \Delta_m^2}}\;.
\end{equation}
The superfluid density is
\begin{equation}
    \rho_s = \omega_p^2 \times  2 \pi T \!\sum_{n >0}^\infty  \frac{\tilde \Delta_n^2}{\left(\tilde \omega_n^2+ \tilde \Delta_n^2 \right)^{3/2}}\;.
\end{equation}

To model the strong-coupling regime of \swave\ superconductivity, the parameters of the boson spectrum are taken to be the same as for the \dwave\ case.

\subsection{Optical conductivity and normal-state scattering rate}
\label{sec:scattering_rate}

The Eliashberg equations can be analytically continued onto the real frequency axis to obtain dynamical properties.  In the absence of vertex corrections, the normal-state optical conductivity can be written
\begin{equation}
    \sigma(\nu) = i\frac{\epsilon_0\omega_p^2}{\nu}\!\!\!\int_{-\infty}^{+\infty}\!\!\!\!\!\!\!d\omega\left[f\left(\omega^\prime\right)-f\left(\omega\right)\right] \frac{1}{\tilde \omega(\omega^\prime) - \tilde \omega^\ast(\omega)}\;,
    \label{eq:normal_state_conductivity}
\end{equation}
where $f(\omega)$ is the Fermi function at temperature $T$ and \mbox{$\omega^\prime = \omega+\nu$}.  Following Ref.~\onlinecite{Schachinger.1997}, the renormalized frequency is given on the real axis by
\begin{equation}
\begin{split}
    \tilde \omega(\omega) & = \omega + i \Gamma_N + i \pi T \sum_{m=0}^\infty \left[\lambda(\omega\!-\!i\omega_m)-\lambda(\omega\!+\!i\omega_m) \right]\\
    & + i \frac{\pi}{2} \int_{-\infty}^\infty d\Omega I^2 F(\Omega) \left[\coth\left(\tfrac{\Omega}{2T}\right) + \tanh\left(\tfrac{\omega-\Omega}{2T}\right)\right]\;,
    \label{eq:renorm_freq_realaxis_scm}
\end{split}
\end{equation}
where we have specialized to the normal state. It is useful to express the $\lambda$ function as
\begin{equation}
    \lambda(\omega) = - \int_{-\infty}^\infty \frac{d\Omega I^2 F(\Omega)}{\omega - \Omega + i \eta}\;,
\end{equation}
so that the Matsubara-sum term in Eq.~\ref{eq:renorm_freq_realaxis_scm} becomes
\begin{equation}
 \pi T\!\!\int_{-\infty}^\infty \!\!\!\!d\Omega I^2 F(\Omega)\sum_{m=0}^\infty \left[\frac{1}{\omega_m + i(\omega - \Omega)} + \frac{1}{\omega_m - i(\omega - \Omega)}\right]\;.
    \label{eq:renorm_freq_Mat_sum}
\end{equation}
For fixed $\Omega$ the summation term diverges, but can be regularized by noting that $I^2F(-\Omega) = - I^2 F(\Omega)$; pairing $\pm \Omega$ terms; and using
\begin{equation}
    \sum_{n=0}^\infty \left[\frac{1}{n+a}-\frac{1}{n+b}\right] = \psi(b) - \psi(a)\;.
\end{equation}
The Matsubara-sum term in Eq.~\ref{eq:renorm_freq_realaxis_scm} then evaluates to
\begin{equation}
    \begin{split}
        \tfrac{1}{2}\!\int_0^\infty d\Omega I^2 F(\Omega)\left[\psi\left(\tfrac{1}{2}+i\tfrac{\omega+\Omega}{2\pi T}\right) + \psi\left(\tfrac{1}{2}-i\tfrac{\omega+\Omega}{2\pi T}\right)\right.\\ 
        - \left.\psi\left(\tfrac{1}{2}+i\tfrac{\omega-\Omega}{2\pi T}\right) -\psi\left(\tfrac{1}{2}-i\tfrac{\omega-\Omega}{2\pi T}\right) \right]\;.
     \label{eq:renorm_freq_Mat_sum_term}
    \end{split}
\end{equation}
For the final term of Eq.~\ref{eq:renorm_freq_realaxis_scm}, we use the identity
\begin{equation}
    i \pi \tanh(\pi x) = \psi\left(\tfrac{1}{2} + i x\right) - \psi\left(\tfrac{1}{2} - i x\right)
\end{equation}
and again group $\pm \Omega$ terms into a single integral over positive $\Omega$ to obtain
\begin{equation}
\begin{split}
     i \frac{\pi}{2} \int_{0}^\infty d\Omega I^2 F(\Omega)\left[\tanh\left(\tfrac{\omega-\Omega}{2T}\right)- \tanh\left(\tfrac{\omega+\Omega}{2T}\right)\right]\\
      = \tfrac{1}{2}\int_{0}^\infty d\Omega I^2 F(\Omega)\left[\psi\left(\tfrac{1}{2}+i\tfrac{\omega-\Omega}{2\pi T}\right) - \psi\left(\tfrac{1}{2}-i\tfrac{\omega-\Omega}{2\pi T}\right)\right.\\ 
        - \left.\psi\left(\tfrac{1}{2}+i\tfrac{\omega+\Omega}{2\pi T}\right) + \psi\left(\tfrac{1}{2}-i\tfrac{\omega+\Omega}{2\pi T}\right) \right]\;.
    \label{eq:renorm_freq_realaxis_term_two}
\end{split}
\end{equation}
Combining Eqs.~\ref{eq:renorm_freq_Mat_sum_term} and \ref{eq:renorm_freq_realaxis_term_two}, we are left with
\begin{equation}
    \begin{split}
        \tilde \omega(\omega) & = \omega + i \Gamma_N + \int_0^\infty d\Omega\,I^2 F(\Omega)\left[ i \pi \coth\left(\tfrac{\Omega}{2T}\right)\right.\\
         & + \left. \psi\left(\tfrac{1}{2}-i\tfrac{\omega+\Omega}{2\pi T}\right) - \psi\left(\tfrac{1}{2}-i\tfrac{\omega-\Omega}{2\pi T}\right)\right]\;,
    \end{split}
    \label{eq:renormized_frequency_LRZ}
\end{equation}
similar to the expression given in Ref.~\onlinecite{Lee.1997}.

After evaluating the conductivity according to Eq.~\ref{eq:normal_state_conductivity}, the normal-state transport scattering rate, $\Gamma_\mathrm{tr}$, including the effects of inelastic scattering, is obtained from the conductivity phase angle in the low frequency limit:
\begin{equation}
    \Gamma_\mathrm{tr} \equiv \lim_{\nu \to 0} \nu \frac{\mathrm{Re}[\sigma(\nu)]}{\mathrm{Im}[\sigma(\nu)]}\;.
\end{equation}

\begin{table*}[t]
\renewcommand{\arraystretch}{1.12}
\begin{tabular}
{|c|c|c|c|c|c|}
\hline
Material & $T_c (\mathrm{K})$ & $\omega_{p,D}~(\mathrm{cm}^{-1})$ & $f_s = \omega_{p,s}^2/\omega_{p,D}^2$ & $k_B T_c/\hbar \Gamma_\mathrm{tr}$ & Remarks \\
\hline
Nb & 8.3 & 74500 \cite{Romaniello.2006} & 0.0558  \cite{Pronin.1998} & 0.0156  \cite{Pronin.1998} &  \\
Nb & 9.3 & 74500 \cite{Romaniello.2006} &  0.231 \cite{Klein.1994} & 0.0593  \cite{Klein.1994}&  \\
Pb & 7.2 & 59400 \cite{Ordal:85} & 0.476 \cite{Klein.1994} & 0.119 \cite{Klein.1994} &  \\
TiN & 3.4 & 23660 \cite{Pfuner.2009} & 0.00849  \cite{Pracht.2012} &  0.00161 \cite{Pracht.2012} &  \\
\hline
Nb-SrTiO$_3$ & 0.168 & 1266 \cite{vanderMarel.2011} & 0.278 \cite{Thiemann.2018} & 0.0753 \cite{Thiemann.2018} & 0.1\% Nb doped \\
Nb-SrTiO$_3$ & 0.254 & 1888 \cite{vanderMarel.2011} & 0.245 \cite{Thiemann.2018} & 0.0521 \cite{Thiemann.2018} & 0.2\% Nb doped \\
Nb-SrTiO$_3$ & 0.346 & 2735 \cite{vanderMarel.2011} & 0.184 \cite{Thiemann.2018} & 0.0371 \cite{Thiemann.2018} & 0.35\% Nb doped \\
Nb-SrTiO$_3$ & 0.278 & 3651 \cite{vanderMarel.2011} & 0.102 \cite{Thiemann.2018} & 0.0207 \cite{Thiemann.2018} & 0.5\% Nb doped \\
Nb-SrTiO$_3$ & 0.213 & 3773 \cite{vanderMarel.2011} & 0.058 \cite{Thiemann.2018} & 0.0111 \cite{Thiemann.2018} & 0.7\% Nb doped \\
\hline
\lco\ & 40 & 5660 \cite{Quijada.1995,Tanner.1998,Liu:1999} & 0.80 \cite{Quijada.1995,Tanner.1998,Liu:1999} & 0.335 \cite{Quijada.1995,Tanner.1998,Liu:1999}&  \\
Tl$_2$Ba$_2$CaCu$_2$O$_{8+\delta}$ & 110 & 10347 \cite{Zibold.1998} & 0.88 \cite{Zibold.1998,Tanner.1998,Liu:1999} & 0.348 \cite{Zibold.1998,Tanner.1998,Liu:1999} &  \\
\bscco\ & 85 & 9040 \cite{Quijada.1999}& 0.95 \cite{Tanner.1998,Quijada.1999} & 0.462 \cite{Quijada.1999} & $a$-axis \\
\bscco\ & 85 & 8750 \cite{Quijada.1999}& 0.94 \cite{Tanner.1998,Quijada.1999} & 0.376 \cite{Quijada.1999} & $b$-axis \\
\bscco\ & 85 & 8800 \cite{Liu:1999} & 0.88 \cite{Liu:1999} & 0.369 \cite{Liu:1999} & \\
\ybco{7} & 91 & 9800 \cite{Tanner.1998} & 0.92 \cite{Tanner.1998} & 0.408 \cite{Tanner.1998} &  \\
\ybco{7-\delta} & 92 & 9200 \cite{Liu:1999} & 0.89 \cite{Liu:1999} & 0.413 \cite{Liu:1999} &  \\
\hline
\lsco\ & 27.5 & 9790 \cite{Mahmood:2017} & 0.423 \cite{Mahmood:2017} & 0.468 \cite{Mahmood:2017} &  \\
\lsco\ & 26.0 & 9790 \cite{Mahmood:2017} & 0.371 \cite{Mahmood:2017} & 0.380 \cite{Mahmood:2017} &  \\
\lsco\ & 19.5 & 9790 \cite{Mahmood:2017} & 0.227 \cite{Mahmood:2017} & 0.266 \cite{Mahmood:2017} &  \\
\lsco\  & 17.5 & 9790 \cite{Mahmood:2017} & 0.206 \cite{Mahmood:2017} & 0.242 \cite{Mahmood:2017} & overdoped  \\
\lsco\  & 13.5 & 9790 \cite{Mahmood:2017} & 0.122 \cite{Mahmood:2017} & 0.161 \cite{Mahmood:2017} &  \\
\lsco\  & 10.5 & 9790 \cite{Mahmood:2017} & 0.065 \cite{Mahmood:2017} & 0.110 \cite{Mahmood:2017} &  \\
\lsco\  & 7.0 & 9790 \cite{Mahmood:2017} & 0.031 \cite{Mahmood:2017} & 0.073 \cite{Mahmood:2017} &  \\
\hline
\ybco{6.333} & 16.9 & 4159 \cite{Padilla:2005ir} & 0.512 \cite{Broun:2007p49} & 0.802 \cite{Broun:2007p49} &  \\
\ybco{6.333} & 16.2 & 4128 \cite{Padilla:2005ir} & 0.482 \cite{Broun:2007p49} & 0.778 \cite{Broun:2007p49} &  \\
\ybco{6.333} & 15.8 & 4112 \cite{Padilla:2005ir} & 0.466 \cite{Broun:2007p49} & 0.739 \cite{Broun:2007p49} &  \\
\ybco{6.333} & 15.2 & 4086 \cite{Padilla:2005ir} & 0.441 \cite{Broun:2007p49} & 0.709 \cite{Broun:2007p49} &  \\
\ybco{6.333} & 14.7 & 4064 \cite{Padilla:2005ir} & 0.421 \cite{Broun:2007p49} & 0.669 \cite{Broun:2007p49} &  \\
\ybco{6.333} & 14.1 & 4037 \cite{Padilla:2005ir} & 0.396 \cite{Broun:2007p49} & 0.637 \cite{Broun:2007p49} &  \\
\ybco{6.333} & 13.1 & 3991 \cite{Padilla:2005ir} & 0.357 \cite{Broun:2007p49} & 0.585 \cite{Broun:2007p49} &  \\
\ybco{6.333} & 12.6 & 3971 \cite{Padilla:2005ir} & 0.339 \cite{Broun:2007p49} & 0.546 \cite{Broun:2007p49} &  \\
\ybco{6.333} & 12.1 & 3946 \cite{Padilla:2005ir} & 0.319 \cite{Broun:2007p49} & 0.491 \cite{Broun:2007p49} & underdoped \\
\ybco{6.333} & 11.6 & 3924 \cite{Padilla:2005ir} & 0.301 \cite{Broun:2007p49} & 0.456 \cite{Broun:2007p49} &  (doping tuned continuously \\
\ybco{6.333} & 11.2 & 3906 \cite{Padilla:2005ir} & 0.287 \cite{Broun:2007p49} & 0.428 \cite{Broun:2007p49} & via CuO-chain oxygen order) \\
\ybco{6.333} & 10.5 & 3872 \cite{Padilla:2005ir} & 0.262 \cite{Broun:2007p49} & 0.382 \cite{Broun:2007p49} &   \\
\ybco{6.333} & 9.9 & 3845 \cite{Padilla:2005ir} & 0.242 \cite{Broun:2007p49} & 0.345 \cite{Broun:2007p49} &  \\
\ybco{6.333} & 9.4 & 3820 \cite{Padilla:2005ir} & 0.224 \cite{Broun:2007p49} & 0.301 \cite{Broun:2007p49} &  \\
\ybco{6.333} & 8.7 & 3791 \cite{Padilla:2005ir} & 0.204 \cite{Broun:2007p49} & 0.265 \cite{Broun:2007p49} &  \\
\ybco{6.333} & 7.9 & 3752 \cite{Padilla:2005ir} & 0.179 \cite{Broun:2007p49} & 0.208 \cite{Broun:2007p49} &  \\
\ybco{6.333} & 7.0 & 3706 \cite{Padilla:2005ir} & 0.151 \cite{Broun:2007p49} & 0.168 \cite{Broun:2007p49} &  \\
\ybco{6.333} & 6.4 & 3680 \cite{Padilla:2005ir} & 0.135 \cite{Broun:2007p49} & 0.148 \cite{Broun:2007p49} &  \\
\ybco{6.333} & 6.0 & 3660 \cite{Padilla:2005ir} & 0.124 \cite{Broun:2007p49} & 0.116 \cite{Broun:2007p49} &  \\
\ybco{6.333} & 4.2 & 3571 \cite{Padilla:2005ir} & 0.079 \cite{Broun:2007p49} & 0.061 \cite{Broun:2007p49} &    \\ 
\hline
\sro\ & 1.5 & 12240 \cite{Tamai.2019,Khasanov.2023} & 0.941 \cite{Landaeta.2024} & 4.17  \cite{Barber.2018} & \\
\cecoin\ & 2.25 & 8660 \cite{Truncik.2013} & 0.879 \cite{Truncik.2013} & 0.346 \cite{Truncik.2013} &  \\
\hline
\end{tabular}
\caption{\label{tab:materials} {Experimental data used in Figs.~\ref{fig:standard_Homes_scaling}} to \ref{fig:modified_Homes_scaling_inelastic}, with references to original sources.  The revised superfluid scaling requires four quantities, \mbox{($\rho_{s0}$, $\omega_{p,D}^2$, $\Gamma_\mathrm{tr}$ and $T_c$)}, used to generate the ratios $f_s = \omega_{p,s}^2/\omega_{p,D}^2$ and $k_B T_c/\hbar \Gamma_\mathrm{tr}$.  Detailed methods are discussed in Sec.~\ref{sec:experimental_data}.}
\end{table*}

\section{Experimental data}

\label{sec:experimental_data}

The experimental data in the figures are drawn from a variety of sources, with a strong emphasis towards optical, or optical-like, data, as with the original Homes scaling paper.  For a material to be placed in the revised superfluid scaling plot, four quantities are required: \mbox{($\rho_{s0}$, $\omega_{p,D}^2$, $\Gamma_\mathrm{tr}$ and $T_c$)}, with $\omega_{p,D}^2$ the crucial new input, not necessary for the original Homes plot.  The data are summarized in Table~\ref{tab:materials} and their sources discussed in this section.  The superconducting transition temperature, $T_c$, is usually the simplest quantity to determine and requires no separate explanation.

\subsection{Superfluid density}

Superfluid density, $\rho_{s0} \equiv \omega_{p,s}^2$, has been of central importance in the study of cuprates and other unconventional superconductors due to its close connection to order-parameter symmetry \cite{HARDY:1993p632}, as well as setting the energy scale for fluctuations of the order-parameter phase \cite{Emery1995}.  It is, however, a difficult quantity to measure absolutely and accurately, with a wide variation in reported results between different techniques, and different samples of the same material.  For this reason the original Homes scaling paper focused on optical measurements of superfluid density, where the reactive superfluid response shows up as a dominant $1/\nu$ term in the imaginary part of the optical conductivity, $\sigma_2(\nu)$, that can be isolated using
\begin{equation}
\epsilon_0 \omega_{p,s}^2 = \lim_{\nu \to 0} \nu \sigma_2(\nu)\;.
\label{eq:superfluid_density_from_sigma2}
\end{equation}
For reflectance measurements on bulk single crystals, which formed the bulk of the data in the original Homes plot, careful Kramers--Kronig analysis over a wide frequency range is required to obtain $\sigma_2(\nu)$, and the measurements must extend low enough in frequency that the $\nu \to 0$ limit of $\nu \sigma_2(\nu)$ is well resolved.  The $\rho_{s0}$ data for Nb \cite{Pronin.1998}, Pb \cite{Klein.1994}, and the near optimally doped cuprates have been obtained in this way \cite{Quijada.1995,Tanner.1998,Liu:1999,Zibold.1998,Quijada.1999}.  THz time-domain spectroscopy (THz-TDS) offers an alternative means of obtaining $\rho_s$ and other quantities and, in the transmission geometry for thin films, the coherent nature of the technique allows the complex conductivity to be measured directly, without Kramers--Kronig analysis.  This has been put to great effect on MBE-grown films of overdoped \lsco\ \cite{Mahmood:2017}.  A variation of the technique (in the frequency domain rather than time domain) was used to obtain the data on TiN \cite{Pracht.2012}, again in a phase-coherent manner.

Other techniques for obtaining superfluid density are usually thought of as more direct measurements of penetration depth, although the microwave surface impedance measurements used for Nb-SrTiO$_3$ \cite{Thiemann.2018}, underdoped \ybco{6.333} \cite{Broun:2007p49} and \cecoin\ \cite{Truncik.2013} are really just low frequency applications of the optical technique and, because both real and imaginary parts of the surface impedance were separately resolved, also act as a phase-coherent measurement.  Finally, superfluid density in a nanofabricated sphere of \sro\ was obtained via measurements of the lower critical field $H_{c1}$ \cite{Landaeta.2024}, a technique that is very reliable in the experimental geometry used, which should eliminate surface-barrier effects.

\subsection{Drude weight}

\begin{figure}[t]
    \centering
        \includegraphics[width=0.75\columnwidth,scale=1.0]{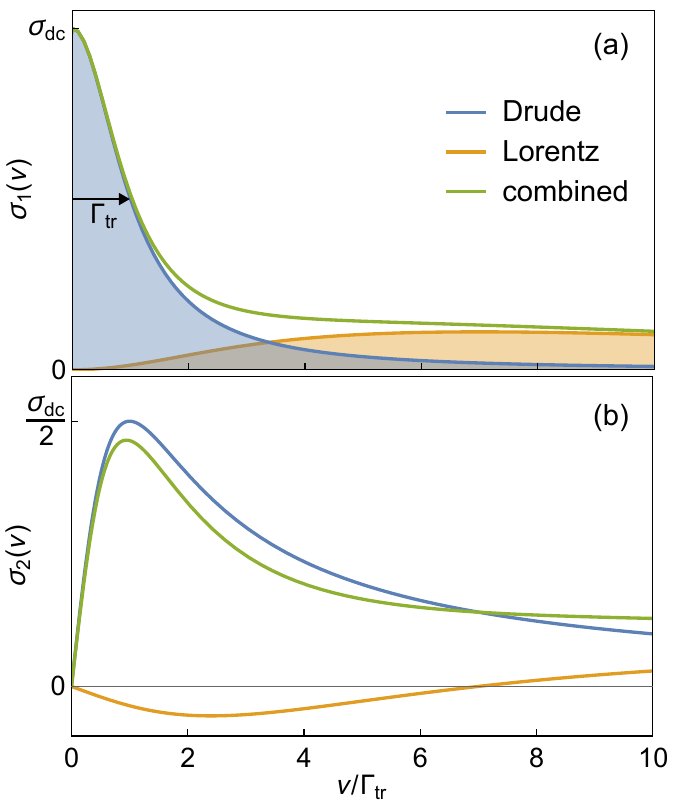}
    \caption{Model normal-state spectrum containing Drude and Lorentz components, showing real and imaginary parts of the normal-state conductivity. (a) Drude--Lorentz fits to $\sigma_1(\nu)$ can be used to isolate the Drude contribution to the integrated weight, shown shaded in blue. (b) The Drude component dominates $\sigma_2(\nu)$ at low frequencies, making complex conductivity, measured by techniques such as THz-TDS, a powerful and unambiguous means of determining the Drude component of the spectrum, and therefore $\omega_{p,D}^2$.}
    \label{fig:DrudeLorentz}
\end{figure}

The idea of Drude weight follows from the optical sum rule,
\begin{equation}
    \int_0^{\nu_c} \sigma_1(\nu) d\nu = \frac{\pi}{2} \epsilon_0 \omega_p^2\;,
\end{equation}
which implies that optical weight can be redistributed in frequency but must otherwise be conserved.  Here the cutoff frequency, $\nu_c$, is chosen to include the conduction electrons while excluding interband transitions and other high energy processes.  In the simplest situation of an isotropic, noninteracting band relaxing at a single rate $\Gamma_\mathrm{tr}$, we have the Drude result:
\begin{equation}
    \sigma_1(\nu) = \frac{n e^2}{m} \frac{\Gamma_\mathrm{tr}}{\nu^2 + \Gamma^2_\mathrm{tr}}\;,
\end{equation}
a Lorentzian spectrum of width $\Gamma_\mathrm{tr}$, whose weight can be integrated to obtain $\omega_{p,D}^2 = \omega_p^2 = ne^2/\epsilon_0 m$.  In the presence of interactions (e.g., with bosonic fluctuations)  the situation is more complicated --- interactions shift part of the spectral weight to higher frequencies, leaving behind a reduced Drude weight near $\nu = 0$. As mentioned above, a particularly clear example of this spectral weight transfer is observed in  \cecoin\ \cite{Singley.2002}, as Kondo physics gradually shifts weight to a singlet--triplet absorption at finite frequency, leaving behind the renormalized heavy-quasiparticle band at low temperature.  In general, the optical spectra of interacting systems can be fit to a Drude-like spectrum plus one or more Lorentz oscillators, as illustrated in Fig.~\ref{fig:DrudeLorentz}(a), and this process has been used to obtain the Drude weights tabulated here for Nb \cite{Romaniello.2006}, Pb \cite{Ordal:85}, Nb-SrTiO$_3$ \cite{vanderMarel.2011}, the near optimally doped cuprates \cite{Quijada.1995,Tanner.1998,Liu:1999,Zibold.1998,Quijada.1999} and for underdoped \ybco{6.333} \cite{Padilla:2005ir}, where an interpolation of the doping-dependent data of Ref.~\onlinecite{Padilla:2005ir} was employed. (For the near optimally doped cuprates, our decision to focus on the optical data of Tanner et al.\ is due to the particularly careful analysis of reflectance data used to isolate $\omega_{p,D}^2$ \cite{Quijada.1995,Tanner.1998,Liu:1999,Zibold.1998,Quijada.1999}.) Ambiguity can arise, and care must be taken, when the Lorentz oscillator starts to overlap with the low frequency Drude peak.  In this situation, transmission THz-TDS on thin films comes into its own, as the phase-coherent measurement of the normal-state complex conductivity allows a direct detemination of $\omega_{p,D}^2$ in the low frequency limit:
\begin{equation}
    \epsilon_0\omega_{p,D}^2 \approx \lim_{\nu \to 0} \nu \frac{\sigma_1^2(\nu)}{\sigma_2(\nu)}\;.
\end{equation}
This use of THz-TDS is particularly powerful, as can be seen in the imaginary part of the conductivity, in Fig.~\ref{fig:DrudeLorentz}(b).  THz-TDS (in the form of fits to the complex conductivity) has been used to obtain $\omega_{p,D}^2$ for overdoped \lsco\ \cite{Mahmood:2017}, with further comments on that material below.  Due to the heavy renormalization of $\Gamma_\mathrm{tr}$, the microwave measurements on \cecoin\ in Ref.~\onlinecite{Truncik.2013} span a sufficiently broad frequency range that $\omega_{p,D}^2$ can be obtained directly from the high frequency, low temperature conductivity, in a  manner similar to Eq.~\ref{eq:superfluid_density_from_sigma2}:
\begin{equation}
\epsilon_0 \omega_{p,D}^2 = \lim_{\nu \gg \Gamma_\mathrm{tr}} \nu \sigma_2(\nu)\;, 
\label{eq:Drude_weight_from_sigma2}
\end{equation}
as shown in Fig.~6c of Ref.~\onlinecite{Truncik.2013}.

\begin{figure*}[t]
    \centering
        \includegraphics[width=0.9\linewidth,scale=1.0]{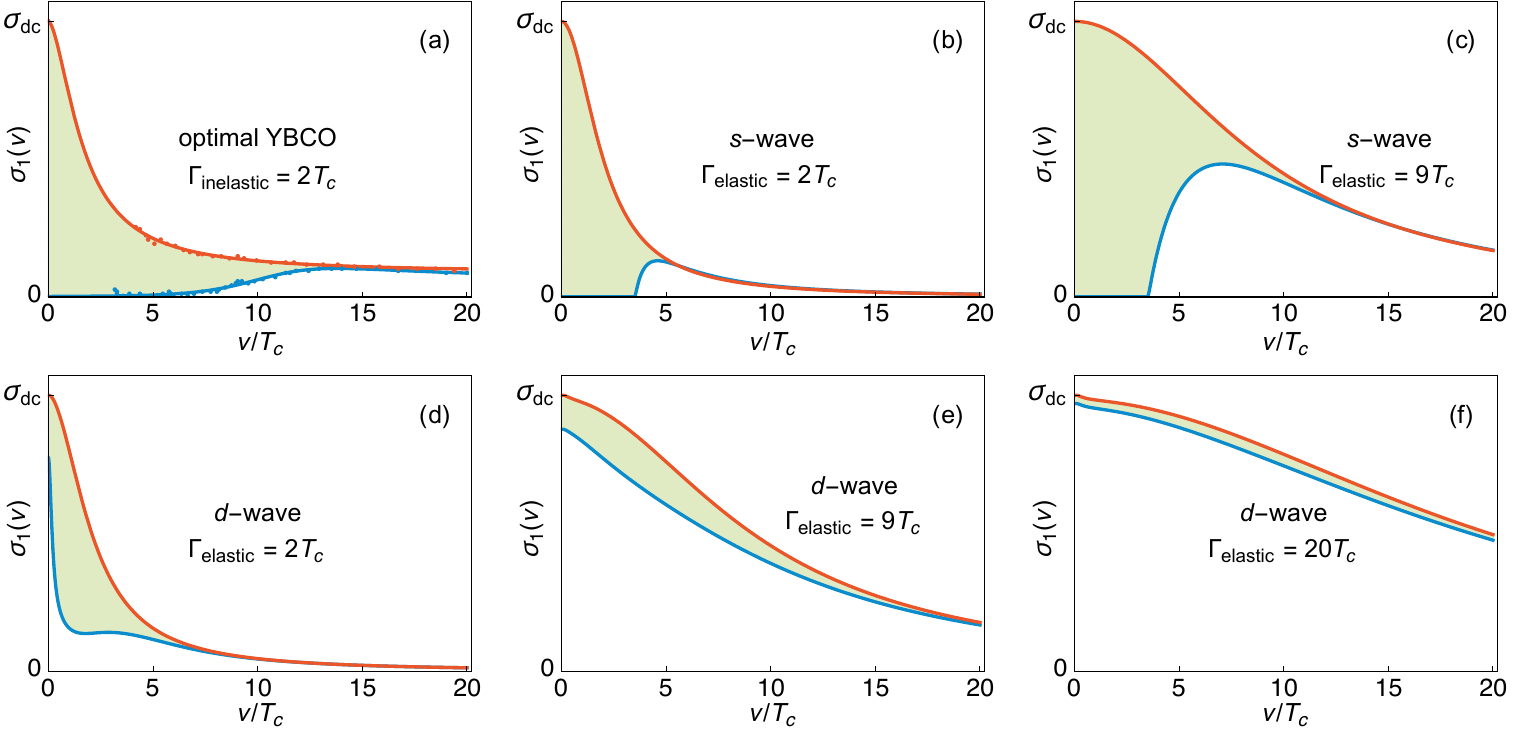}
    \caption{Optical conductivity spectra, $\sigma_1(\nu)$, for various superconductors, illustrating the formation of the zero-temperature superfluid from the normal-state Drude weight.  Each plot contains a normal-state spectrum ($T > T_c$) and a superconducting spectrum in the $T \to 0$ limit, with the shaded area in between indicating the spectral weight that condenses to form the zero-temperature superfluid density. (a) The optical conductivity of optimally doped \ybco{7-\delta}, based on optical data from Ref.~\onlinecite{Rotter1991} and dc conductivity from Ref.~\onlinecite{Ando2004}, showing Planckian-limit inelastic scattering ($\Gamma_\mathrm{tr} \approx 2 T_c$) in the normal state. (b) and (c) Optical conductivity of a dirty \swave\ superconductor, for normal-state elastic scattering rates $\Gamma_\mathrm{tr} = 2 T_c$ and $9 T_c$, based on the Zimmermann's calculation \cite{Zimmermann1991}. (d) to (f) Optical conductivity of a dirty \dwave\ superconductor, calculated for normal-state elastic scattering rates $\Gamma_\mathrm{tr} = 2 T_c$, $9 T_c$ and $20 T_c$, using an impurity model and formalism taken from {\it ab initio} calculations for overdoped \lsco\ \cite{Ozdemir:2022,Broun.2024}. 
}
    \label{fig:ConductivitySpectra}
\end{figure*}

Finally, the Drude weight can also be calculated from band structure, as long as the energy dispersion is  known to sufficiently low energies that interaction-induced band renormalization is included.  This has been done for the case of \sro\ \cite{Khasanov.2023}, based on high resolution laser ARPES \cite{Tamai.2019}.  It has also informed our decision to use a doping-independent value for $\omega_{p,D}^2$ in overdoped \lsco, where both calculations \cite{Lee-Hone:2018} based on ARPES band structure \cite{Yoshida:2006hw}, and optical measurements  for \lcco\ over a wide doping range \cite{Kim2021}, reveal negligible doping dependence of $\omega_{p,D}^2$ in the overdoped regime.  We have therefore set $\omega_{p,D}^2$ for overdoped \lsco\ to the average of the values measured for the $T_c = 17.5$, 19.5, 26 and 27.5~K samples in Fig.~S7 of Ref.~\onlinecite{Mahmood:2017}.

\subsection{Transport scattering rate}

Optical measurements allow the normal-state transport scattering rate to be determined directly as the width of the Drude component of the optical spectrum, usually on the basis of a Drude fit to $\sigma(\nu)$.  This has been used to obtain $\Gamma_\mathrm{tr}$ in the case of the near optimally doped cuprates \cite{Quijada.1995,Tanner.1998,Liu:1999,Zibold.1998,Quijada.1999} and the THz measurements on overdoped \lsco\ \cite{Mahmood:2017}. 

In all other cases, $\Gamma_\mathrm{tr}$ has been inferred from the dc conductivity \cite{Klein.1994,Broun:2007p49,Truncik.2013} or resistivity \cite{Pronin.1998,Pracht.2012,Thiemann.2018,Barber.2018}, using the separately determined value of $\omega_{p,D}^2$ and the Drude result, 
\begin{equation}
    \Gamma_\mathrm{tr} = \epsilon_0 \omega_{p,D}^2/\sigma_\mathrm{dc} = \epsilon_0 \omega_{p,D}^2 \rho_\mathrm{dc}.
\end{equation}
We note that particular care was required in the case of underdoped \ybco{6.333} \cite{Broun:2007p49}, where a pronounced fluctuation peak appears in the 2.64~GHz microwave conductivity, $\sigma_1(T)$, near $T_c$  (see Fig.~5 of Ref.~\onlinecite{Broun:2007p49}).  An interpolation through the $\sigma_1(T)$ data on either side of the fluctuation peak has been performed at each doping, with the interpolation evaluated 2~K above the maximum of the fluctuation peak, in order to estimate $\sigma_\mathrm{dc}(T_c)$ in the absence of superconducting fluctuations.  

\section{Optics and pair breaking}

\label{sec:optics_pair_breaking}

Optical spectra also provide insight into the condensation of spectral weight to form the zero-temperature superfluid density, illustrating why materials with strong inelastic scattering show Homes scaling, and why dirty \swave\ and \dwave\ superconductors follow distinct scaling relations.
In each case, the total spectral weight available for condensation is set by the normal-state conductivity spectrum, which has height equal to the dc conductivity, $\sigma_\mathrm{dc}$, and characteristic width given by the normal-state transport scattering rate, $\Gamma_\mathrm{tr}$.  As shown in Fig.~\ref{fig:DrudeLorentz}, this sets the Drude weight, with $\omega_{p,D}^2 \propto \sigma_\mathrm{dc} \Gamma_\mathrm{tr}$.  As defined earlier, a fraction $f_s = \omega_{p,s}^2/\omega_{p,D}^2$ condenses into the $T \to 0$ superfluid density.  

Three different cases are illustrated in Fig.~\ref{fig:ConductivitySpectra}, starting with experimental data for optimally doped \ybco{7 - \delta} in Fig.~\ref{fig:ConductivitySpectra}(a), based on Refs.~\onlinecite{Rotter1991} and \onlinecite{Ando2004}.  Similar to other optimally doped cuprates, \ybco{7 - \delta} has a normal-state inelastic Drude width of $\Gamma_\mathrm{tr} \approx 2 T_c$ that is characteristic of the Planckian regime, so that the available Drude weight scales as $\sigma_\mathrm{dc} \Gamma_\mathrm{tr} \sim \sigma_\mathrm{dc} T_c$.  The superconducting condensation draws in the majority of the available spectral weight up to $\nu = 10~T_c$, and therefore $\rho_s \propto \sigma_\mathrm{dc} T_c$, in accord with Homes scaling. The key point is that unlike elastic disorder scattering in a \dwave\ superconductor, inelastic scattering does not induce significant low frequency absorption in $\sigma_1(\nu)$ in the superconducting state. The onset of absorption above $6~T_c$ is not a gap edge, but the gradual resumption of inelastic scattering at high energy \cite{Hirschfeld:1996}.

The case of an \swave\ superconductor with purely elastic scattering is plotted in Figs.~\ref{fig:ConductivitySpectra}(b) and (c), calculated using Zimmermann's extenstion \cite{Zimmermann1991} of Mattis--Bardeen theory \cite{Mattis.1958}. At $T = 0$, there is no absorption below the BCS gap edge at $2 \Delta \approx 3.5~k_B T_c$: the spectral weight that condenses into the superfluid again scales as $\sigma_\mathrm{dc} T_c$, illustrating how the dirty \swave\ superconductor mimics Homes scaling, as previously noted \cite{Homes.2005,Kogan:2013eh}.

For the \dwave\ superconductor with elastic-scattering disorder, optical spectra have been calculated using the methods described in Ref.~\onlinecite{Broun.2024}, based on a model of realistic impurity potentials in overdoped \lsco\ with varying concentration of apical oxygen vacancies.  Figures~\ref{fig:ConductivitySpectra}(d) to (f) show spectra for $\Gamma_\mathrm{tr} = 2T_c$, $9T_c$ and $20T_c$.  Note that in a \dwave\ superconductor, there is nothing to protect the low-lying states from pair breaking: as a result, there is absorption at all frequencies and no clear gap edge, with the superfluid drawn from a shrinking sliver of spectral weight as disorder increases.  The ``filling in" of the uncondensed $T \to 0$ spectral weight means that for the disorder-dominated \dwave\ superconductor, the condensed superfluid  scales as $\sigma_\mathrm{dc} T_c \times T_c/\Gamma_\mathrm{tr} \propto (T_c/\Gamma_\mathrm{tr})^2$, resulting in scaling behavior distinct from both the Planckian regime and the dirty \swave\ superconductor.

\section{Discussion}
\label{sec:Discussion}

The original Homes scaling plot provoked considerable excitement when it was published, not only because it exhibited an apparently universal behavior
in those cuprates studied intensively to that time, but because it appeared to establish a bridge between the cuprates and the well-understood conventional $s$-wave superconductors.   In fact much of the variation along the Homes scaling curve was due to the  fact that  both variables ($\rho_s$ and $\sigma_{\mathrm{dc}} T_c$) plotted are proportional to the Drude weight, parameterized by $\omega_{p, D}^2$, which varies substantially from material to material, depending on factors such as band structure, carrier density and mass renormalization.

We believe that the fundamental universal behavior emerges when one removes the $\omega_{p, D}^2$ of both ordinate and abcissa on the Homes plot. The natural dimensionless variables for this revised superfluid scaling plot are then the superfluid fraction, $f_s=\omega_{p, s}^2 / \omega_{p, D}^2$, and the ratio $k_B T_c / \hbar \Gamma_{\mathrm{tr}}$.
A material's position in the plot is then determined by a combination of its order-parameter symmetry and the strength of its elastic and inelastic scattering rates (at $T_c$), relative to $T_c$.  

We show that the revised superfluid scaling plot has four characteristic regimes:
\begin{enumerate}
\item  The strong inelastic scaling regime (or Planckian regime), characterized by a scattering rate at $T_c$ near the Planckian bound, $\hbar \Gamma_{\text {tr }} \sim 2 k_B T_c$, and almost complete condensation of the {\it available} Drude weight into the superfluid condensate, recalling that inelastic scattering has already suppressed the Drude weight significantly relative to the noninteracting system. (As an aside, we note that in optimally doped cuprates this renormalization is approximately a factor of 4, a result  that earlier has been dubbed  ``Tanner's law" \cite{Tanner.1998,Zaanen2004}.) The heavy fermion superconductor \cecoin\ also falls into this regime.  
Within the Migdal--Eliashberg approach, the strong suppression of the normal state Drude weight is associated with 
quite low values of the characteristic boson frequency relative to $T_c$. The strong-coupling regime in many real materials corresponds to boson frequency $\Omega_B \simeq 3.5 T_c$.
\item  The superclean limit, in which $f_s \sim 1$ and $\hbar \Gamma_{\mathrm{tr}} \ll k_B T_c$, indicating that both elastic and inelastic scattering are weak at $T=T_c$. These will typically be low $T_c$ superconductors of exceptional purity, of which \sro\ is so far our best example.  It would be interesting to confirm our analysis  by finding examples of ultraclean $s$-wave superconductors that scale in a similar way.
\item Dirty $s$-wave scaling, radiating out from the Planckian regime as $f_s \propto T_c / \Gamma_{\mathrm{tr}}$, where elastic scattering dominates and $\hbar \Gamma_{\mathrm{tr}} \gg k_B T_c$. Most conventional superconductors fall into this regime, with Pb, the classic strong-coupling $s$-wave material, showing clear signs of inelastic effects \cite{Mori.2008}.
\item Dirty $d$-wave scaling, extending from the Planckian regime as $f_s \propto\left(T_c / \Gamma_{\mathrm{tr}}\right)^2$ as the system becomes dominanted by elastic scattering disorder. As expected from our recent body of work on dirty $d$-wave superconductivity \cite{Lee-Hone:2017,Lee-Hone:2018,LeeHone2020,Ozdemir_comment,Ozdemir:2022,Broun.2024}, this is typified by overdoped $\mathrm{La}_{2-x} \mathrm{Sr}_x \mathrm{CuO}_4$, for which excellent THz measurements have been made on a series of high quality, MBE-grown films with $T_c$ ranging from 7 to 27~K \cite{Mahmood:2017}.
    \end{enumerate}
 Our analysis is built entirely on the BCS--Landau paradigm of well-defined quasiparticles and their transport
    properties.  Surprisingly, and perhaps most importantly for the cuprate phase diagram as a whole, heavily underdoped
\ybco{6.333} \cite{Broun:2007p49}, with $T_c$ ranging from 4 to 17~K , also follows the dirty \dwave\ line. This is a strong indication that Fermi liquid theory  remains operative in the underdoped regime despite  the presence of the pseudogap and  various interwined orders.  It appears that for underdoped cuprates {\it away} from the Planckian regime, there exists an itinerant low-energy charge sector (e.g., nodal arcs) that is still well described as a renormalized Fermi liquid.   In this understanding of the phase diagram, the fact that underdoped \ybco{6.333} begins to peel away from the weak-coupling $d$-wave regime, toward the strong coupling regime, as doping increases may be understood by the decrease in $\Omega_B/T_c$ as $T_c$ rises. This can be seen to a lesser extent in \lsco, but both materials must eventually hook smoothly back into the Planckian regime at optimal doping.  Running this argument in reverse,  we are seeing how disorder-induced $T_c$ suppression turns a strong-coupling superconductor into a weak-coupling superconductor.

\section{Conclusions}
\label{sec:Conclusions}

We have argued that what is generally referred to as Homes scaling, applying to cuprates and a variety of conventional superconductors, hides a more subtle and interesting universality.   This ``revised superfluid scaling", which relates the superfluid fraction to the dimensionless dc conductivity, appears when the Drude weight is removed from the scaling variables and  reflects sensitively both the symmetry  of the superconducting order and the relative size of elastic and inelastic scattering.   Using optical data where available, we have exhibited the revised form of scaling for 45 different materials including cuprates, conventional $s$-wave and heavy fermion superconductors.

In addition to providing a much clearer understanding of the underlying origin of Homes scaling (and observed departures from Homes scaling!), the revised superfluid scaling plot should serve several important purposes in future studies.
First, the models and assumptions used here can be tested and extended by the addition of new data to the plot, with the stringent requirement that careful measurements of Drude weight be carried out as part of that process, particularly in materials in which Drude weight is a sensitive function of material composition, such as cuprates. Coherent THz time-domain spectroscopy on high quality thin films provides one of the best techniques for probing all the necessary quantities $(\rho_{s0}, \omega_{p, D}^2, \Gamma_{\mathrm{tr}}$ and $T_c)$ at the same time and on the same sample.

Secondly, the location of a material within the revised superfluid scaling plot provides immediate insight into its disorder level; the nature of its order parameter; and the strength of coupling to the bosonic fluctuations responsible for superconducting pairing, inelastic scattering and mass renormalization. These are all key properties for understanding a given superconductor, making the revised superfluid scaling plot a promising tool for classifying new materials in the future.

\begin{acknowledgments}
We are grateful for useful discussions with C.~C.~Homes and  D.~B.~Tanner.  D.M.B.\  acknowledges financial support from the Natural Science and Engineering Research Council of Canada.  P.J.H.\ acknowledges support from NSF-DMR-1849751.
\end{acknowledgments}



%

\end{document}